\PassOptionsToPackage{svgnames,dvipsnames}{xcolor}
\documentclass[twocolumn]{aastex701}
\usepackage{CJK}

\usepackage{xcolor}
\definecolor{linkcolor}{cmyk}{1,1,0,0}
\definecolor{lightcoral}{HTML}{F08080}
\definecolor{lightgray}{HTML}{D3D3D3}
\definecolor{matplotlib_green}{HTML}{008000}
\definecolor{matplotlib_orange}{rgb}{1, 0.5, 0.05}
\definecolor{matplotlib_blue}{rgb}{0.12, 0.47, 0.71}
\definecolor{matplotlib_green2}{rgb}{0.18, 0.63, 0.18}
\definecolor{S0}{rgb}{0.78, 0.69, 0.86}
\definecolor{E}{rgb}{0.99, 0.73, 0.51}
\definecolor{S}{rgb}{0.57, 0.80, 0.57}
\definecolor{elliptical}{rgb}{0.57, 0.80, 0.57}
\definecolor{spiral}{rgb}{0.57, 0.80, 0.57}
\definecolor{matplotlib_red}{rgb}{0.84, 0.18, 0.18}

\newcommand{\indep}{\perp \!\!\! \perp}

\usepackage{color,soul}
\usepackage{scalerel}
\usepackage{dashrule}
\usepackage{totcount}
\newtotcounter{citenum}
\def\oldcite{}
\let\oldcite=\bibcite
\def\bibcite{\stepcounter{citenum}\oldcite}
\usepackage{mathtools}
\usepackage[thinc]{esdiff}
\usepackage{fontawesome}
\usepackage{amsmath}
\usepackage{multirow}
\usepackage{fixfoot}
\DeclareFixedFootnote{\repnote}{\url{https://oeis.org/A003024}}
\defcitealias{Jin:2025}{Paper~I}

\received{November 10, 2025}
\revised{February 3, 2026}
\accepted{February 17, 2026}
\submitjournal{\textit{The Astrophysical Journal}}

\shorttitle{Causal Reversal in the $M_\bullet$--$\sigma_0$ Relation}
\shortauthors{Davis et al.}

\begin{document}
\begin{CJK*}{UTF8}{gbsn}

\title{Causal Reversal in the $M_\bullet$--$\sigma_0$ Relation:\\
Implications for High-Redshift Supermassive Black Hole Mass Estimates}

\author[0000-0002-4306-5950,gname=Benjamin,sname=Davis]{Benjamin L.\ Davis}
\affiliation{Center for Astrophysics and Space Science (CASS), New York University Abu Dhabi, PO Box 129188, Abu Dhabi, UAE}
\affiliation{New York University Abu Dhabi, PO Box 129188, Saadiyat Island, Abu Dhabi, UAE}
\email[show]{\href{mailto:ben.davis@nyu.edu}{ben.davis@nyu.edu}}

\author[gname=Saakshi,sname=More]{Saakshi More}
\affiliation{Center for Astrophysics and Space Science (CASS), New York University Abu Dhabi, PO Box 129188, Abu Dhabi, UAE}
\affiliation{New York University Abu Dhabi, PO Box 129188, Saadiyat Island, Abu Dhabi, UAE}
\email{sm9890@nyu.edu}

\author[0009-0000-2506-6645,gname=Zehao,sname=Jin]{Zehao Jin (金泽灏)}
\affiliation{Center for Astronomy and Astrophysics and Department of Physics, Fudan University, Shanghai 200438, People's Republic of China}
\affiliation{Center for Astrophysics and Space Science (CASS), New York University Abu Dhabi, PO Box 129188, Abu Dhabi, UAE}
\affiliation{New York University Abu Dhabi, PO Box 129188, Saadiyat Island, Abu Dhabi, UAE}
\email[show]{\href{mailto:zehaojin@fudan.edu.cn}{zehaojin@fudan.edu.cn}}

\author[0000-0003-3784-5245,gname=Mario,sname=Pasquato]{Mario Pasquato}
\affiliation{Istituto Nazionale di Astrofisica, IASF-Milano Via Alfonso Corti 12, I-20133 Milano, Italy}
\email{mario.pasquato@inaf.it}

\author[0000-0002-8171-6507,gname=Andrea,sname=Macci\`{o}]{Andrea Valerio Macci\`{o}}
\affiliation{Center for Astrophysics and Space Science (CASS), New York University Abu Dhabi, PO Box 129188, Abu Dhabi, UAE}
\affiliation{New York University Abu Dhabi, PO Box 129188, Saadiyat Island, Abu Dhabi, UAE}
\email{maccio@nyu.edu}

\author[0000-0003-3564-6437,gname=Feng,sname=Yuan]{Feng Yuan (袁峰)}
\affiliation{Center for Astronomy and Astrophysics and Department of Physics, Fudan University, Shanghai 200438, People's Republic of China}
\email{\href{mailto:fyuan@fudan.edu.cn}{fyuan@fudan.edu.cn}}

\begin{abstract}

The nascent methodology of applying the principles of causal discovery to astrophysical data has produced affirming results about deeply held theories concerning the causal nature behind the observed coevolution of supermassive black holes (SMBHs) with their host galaxies.
The key results from observations have demonstrated an apparent causal reversal across different galaxy morphologies---SMBHs causally influence the evolution of the physical parameters of their spiral galaxy hosts, whereas SMBHs in elliptical galaxies are passive companions that grow in near lockstep with their hosts.
To further explore and ascertain insights, it is necessary to utilize galaxy simulations to track the time evolution of the observed causal relations to learn more about the temporal nature of the changing SMBH/galaxy evolutionary directions.
We conducted experiments with the NIHAO suite of cosmological zoom-in hydrodynamical simulations to follow the evolution of individual galaxies along with their central SMBH masses ($M_\bullet$) and properties, including central stellar velocity dispersion ($\sigma_0$).
We reproduce the causal results from real galaxies, but add clarity by observing that the SMBH/galaxy causal directions are noticeably inverted between the epochs before and after the peak of star formation.
The implications for causal reversal of the $M_\bullet$--$\sigma_0$ relation portend larger concerns about the reliability of SMBH masses estimated at high redshifts and presumptions of overmassive black holes at early epochs.
Toward this problem, we apply updated causally-informed scaling relations that predict high-$z$ black hole masses that are approximately two orders of magnitude less massive, and thus not overmassive with respect to local $z=0$ SMBH--galaxy mass ratios.

\end{abstract}

\keywords{
\href{http://astrothesaurus.org/uat/159}{Black hole physics (159)};
\href{http://astrothesaurus.org/uat/573}{Galaxies (573)};
\href{http://astrothesaurus.org/uat/591}{Galaxy dynamics (591)};
\href{http://astrothesaurus.org/uat/594}{Galaxy evolution (594)};
\href{http://astrothesaurus.org/uat/602}{Galaxy kinematics (602)};
\href{http://astrothesaurus.org/uat/609}{Galaxy nuclei (609)};
\href{http://astrothesaurus.org/uat/612}{Galaxy physics (612)};
\href{http://astrothesaurus.org/uat/615}{Galaxy properties (615)};
\href{https://astrothesaurus.org/uat/767}{Hydrodynamical simulations (767)};
\href{http://astrothesaurus.org/uat/1663}{Supermassive black holes (1663)}
}

\section{Introduction}

It is widely believed that some fraction of the bolometric luminosity of an accreting supermassive black hole (SMBH) couples thermally with the surrounding gas in its host galaxy.
This energy deposition suggests a feedback mechanism that may regulate the growth of an SMBH by expelling gas from its host galaxy and quenching star formation \citep{Mastichiadis:1995,Ciotti:1997,Ciotti:2001,Silk:1998,Ikeuchi:1999,Renzini:2000,Wyithe:2003}.
Indeed, \citet{Springel:2005} presented hydrodynamical simulations of galaxy mergers, showing how active galactic nuclei (AGNs) feedback regulates star formation and SMBH growth.
The feedback quenches star formation, leading to the formation of red elliptical galaxies, providing a compelling explanation for the observed galaxy color bimodality.
This feedback mechanism suggests a mechanism whereby an SMBH may causally affect its much larger host galaxy, which is typically between $\approx$$10^2$--$10^4$ times more massive at $z=0$ \citep[][their Fig.~12]{Sahu:2019}.

The early theoretical and observational groundwork for AGN feedback was laid three decades ago, and numerical simulations for the past two decades have demonstrated a clear dependency on including the feedback prescriptions in their models in order to reproduce the observed population of red elliptical galaxies in the local Universe.
Along the way, some observational evidence has been presented to support the notion of AGN feedback \citep[e.g.,][]{Schawinski:2007,Fabian:2012,Morganti:2017,Harrison:2024}.
Of course, one of the most notable connections between galaxy formation and SMBHs is the correlation seen between the stellar velocity dispersion of bulges and the masses of SMBHs they host \citep[e.g.,][]{Ferrarese:2000,Gebhardt:2000,Tremaine:2002,Kormendy:2013,Sahu:2019b}.
However, despite the extensive foundation for AGN feedback established by theory/simulations and supportive correlations uncovered from observational searches, there had not been a conclusive affirmation of a causal relation (i.e., a clear cause/effect relationship beyond a superficial correlation), nor its direction (i.e., which observable is a cause and which is an effect).
Recently, studies have sought to bring causal discovery \citep{Pearl:2000,Zanga:2023,Huber:2024} techniques to this problem \citep{Pasquato:2023,Pasquato:2024,Jin:2024,Jin:2025AAS,Jin:2025}.

We studied a sample of 101 SMBHs with dynamically-measured masses in our previous work by \citet{Jin:2025}, hereafter \citetalias{Jin:2025}.
In addition to SMBH mass ($M_\bullet$), their sample also included six measurements of the 101 host galaxies, divided into 35 elliptical, 38 lenticular, and 28 spiral galaxies.
These galaxy measurements included the central stellar velocity dispersion ($\sigma_0$), effective (half-light) radius of the bulge ($R_e$), the average projected density within $R_e$ ($\langle\Sigma_\mathrm{e}\rangle$), total stellar mass of the entire galaxy ($M^*$), color (W2$-$W3), and specific star formation rate (sSFR).
These chosen galaxy parameters provide a well-defined manifold from which to test the causal relationships between the SMBH mass.
While our observational study was not afforded with time series data of the galaxies, the separation of galaxies into their morphologies allowed for a progressive evolutionary study of causal directions from young spiral galaxies to old elliptical galaxies.
In our present study, we repeat the methodology of \citetalias{Jin:2025}, but with simulations to directly track the causal structures with time.

\citetalias{Jin:2025} performed an exhaustive Bayesian analysis of all 1,138,779,265 possible directed acyclic graphs (DAGs) for the possible permutations of their $M_\bullet$ plus six galaxy parameters for morphologically-distinct samples.\footnote{We look up the number of DAGs with $n$ labeled nodes from The On-Line Encyclopedia of Integer Sequences (\url{https://oeis.org/A003024}).}
Among the score-based causal directions uncovered between all variables, perhaps the most revealing result concerns the morphologically-dependent causal directions uncovered in the $M_\bullet$--$\sigma_0$ relation.
\citetalias{Jin:2025} found that in \emph{spiral galaxies}, $\sigma_0\rightarrow M_\bullet$ (i.e., $\sigma_0$ directly causes $M_\bullet$, such that $\sigma_0$ is considered a parent of $M_\bullet$) 22\% of the time.
Conversely for \emph{elliptical galaxies}, $\sigma_0\rightarrow M_\bullet$ in 78\% of all cases.
Unsurprisingly, lenticulars lie somewhere in the middle, with $\sigma_0\rightarrow M_\bullet$ satisfying 72\% of all scenarios.

In \emph{elliptical galaxies}, $\sigma_0$ \emph{is} the causal driver of change (i.e., growth) in $M_\bullet$.
Whereas, $\sigma_0$ \emph{is not} the driver of change in $M_\bullet$ in spiral galaxies, therefore it is more likely that $M_\bullet$ drives the change in $\sigma_0$.
\citetalias{Jin:2025} concluded that this naturally implies that an SMBH is able to affect change in its host spiral galaxy, likely through AGN feedback.
Given the rich supplies of gas typically available in spiral galaxies, this allows a causal pathway for an active SMBH to progressively quench its host galaxy.
By the time a host galaxy has evolved into an elliptical galaxy, the quenching process is complete and the gas supply has been largely exhausted, leaving no medium for an SMBH to continue causally affecting its host galaxy.
This transition places an SMBH as a passive passenger in its host elliptical galaxy, leaving it to only evolve as its host causally allows through accretion and mergers.

The novel work of \citetalias{Jin:2025} offered a unique observational confirmation of AGNs feedback and cemented SMBHs and their host galaxies together via their shared coevolution.
In the present study, we endeavor to advance the study of \citetalias{Jin:2025} into the temporal domain by tracking the time series changes in causal directions between SMBHs and their host galaxies via hydrodynamical simulations.
In particular, our simulations include prescriptions of AGNs feedback.
Thus, the efforts of this work will aim to strengthen the observational confirmation of AGNs feedback presented in \citetalias{Jin:2025}.
Over the course of this paper, we present our methodology of using numerical simulations (\S\ref{sec:theory}), exhibit our results (\S\ref{sec:results}), derive new causally-informed scaling relations (\S\ref{sec:scaling}), and provide a discussion of our findings, their implications, and overall importance (\S\ref{sec:discussion}).
Throughout this work, all mean values are quoted with $\pm1$ standard deviation uncertainties and all median values are quoted with $\pm1$ median absolute deviation uncertainties.

\section{Numerical Simulations}\label{sec:theory}

We conduct our study using simulated galaxies from the NIHAO (Numerical Investigation of a Hundred Astrophysical Objects) Project \citep{Wang:2015,Blank:2019}, which are cosmological zoom-in hydrodynamical simulations performed using the \texttt{Gasoline2} code \citep{Wadsley:2017} with an improved smoothed particle hydrodynamics algorithm.
NIHAO simulations are designed to study galaxy formation and evolution, covering a broad range of dark matter halo masses, from dwarf galaxies to massive ellipticals.
The primary goal of the project is to realistically model the complex processes involved in how galaxies form stars, grow, and change over cosmic time.
NIHAO simulations have been very successful in reproducing observed relationships, such as the inefficiency of galaxy formation (the stellar mass to halo mass relation) and the connection between star formation rates and stellar masses \citep{Blank:2019}.
A key focus is understanding the impact of stellar feedback and the influence of SMBHs on their host galaxies.
By comparing these detailed simulations with observational data, the NIHAO project provides crucial insights into the physical mechanisms that drive galaxy evolution and shape the properties of galaxies, including their dark matter halos, across the Universe.

NIHAO uses a flat $\Lambda$CDM cosmology with parameters from \hspace{-1mm}\citet{Planck:2014}.
The NIHAO simulations span gas particle masses of approximately $6\times10^{2}$--$3\times10^{5}\,\textrm{M}_\odot$ and dark matter particle masses of $3\times10^{3}$--$2\times10^{6}\,\textrm{M}_\odot$, with Plummer-equivalent gravitational softenings ranging from $\sim$50 to $\sim$1000\,pc, depending on halo mass \citep{Wang:2015}.
The NIHAO zoom-in regions are drawn from dark-matter-only parent simulations with comoving box sizes of 15, 20, and 60\,$h^{-1}$\,Mpc, corresponding to volumes of $3.4\times10^{3}$, $8.0\times10^{3}$, and $2.16\times10^{5}$\,$(h^{-1}\,\mathrm{Mpc})^{3}$, respectively \citep{Dutton:2014}.
Halos and subhalos in the NIHAO simulations are identified using the AMIGA Halo Finder \citep[AHF;][]{Knollmann:2009}, and their assembly histories are traced with the AHF MergerTree algorithm based on particle-ID matching across simulation snapshots \citep{Gill:2004}, as described in the NIHAO project \citep{Wang:2015}.
Black holes are seeded in halos above $5\times10^{10}\,\textrm{M}_\odot$ with an initial mass of $10^5\,\textrm{M}_\odot$, grow via a Bondi--Hoyle--Lyttleton accretion model \citep{Hoyle:1939,Bondi:1952} capped at the Eddington limit \citep{Eddington:1916,Eddington:1920}, and inject thermal feedback into the surrounding gas using a subgrid prescription calibrated to reproduce observed black hole and galaxy scaling relations \citep{Blank:2019}.

\subsection{Galaxy Property Extraction}

For each central galaxy in our NIHAO sample, we extract five physical properties for causal discovery analysis, measured at every simulation snapshot with the \texttt{pynbody} package \citep{pynbody}:

\begin{itemize}

\item \textbf{Black hole mass ($M_\bullet$)}: Mass of the \emph{most massive black hole particle} in the central halo, isolating the primary SMBH and excluding lower-mass merger remnants.
Reported in solar units using the \texttt{pynbody} particle filter and halo catalog.

\item \textbf{Stellar mass ($M^\ast$)}: Total mass of all bound star particles in the central halo, converted to solar mass units.

\item \textbf{Effective radius ($R_e$)}: Three-dimensional spherical half-light radius.
Defined as the radius enclosing half of the total $V$-band luminosity of the stellar population, reported in kiloparsecs.

\item \textbf{Central velocity dispersion ($\sigma_0$)}: Stellar velocity dispersion of bulge stars within 5\,kpc and with negative angular momentum (random, non-rotational motion) along the $z$-axis ($j_z < 0$).
A logarithmic radial profile (100 bins) is constructed, and the dispersion is interpolated within 595\,pc.\footnote{Chosen to match the homogenized dispersions of \citet{Jorgensen:1995} in HyperLEDA \citep{hyperleda} and the observational analysis of \citetalias{Jin:2025}.}

\item \textbf{Specific star formation rate (sSFR)}: Ratio of the star formation rate (SFR) to total stellar mass.
The SFR is obtained by binning the star formation history into the 64 simulation snapshots (216\,Myr spacing).
The resulting sSFR, in units of yr$^{-1}$, quantifies star formation activity relative to stellar mass since the last snapshot.
\end{itemize}

\end{CJK*}

\section{Results}\label{sec:results}

\subsection{Star-forming and Quenched Galaxies}\label{sec:SF}

\begin{figure*}
    \centering
    \includegraphics[clip=true, trim= 2mm 2mm 0mm 0mm, width=0.49\linewidth]{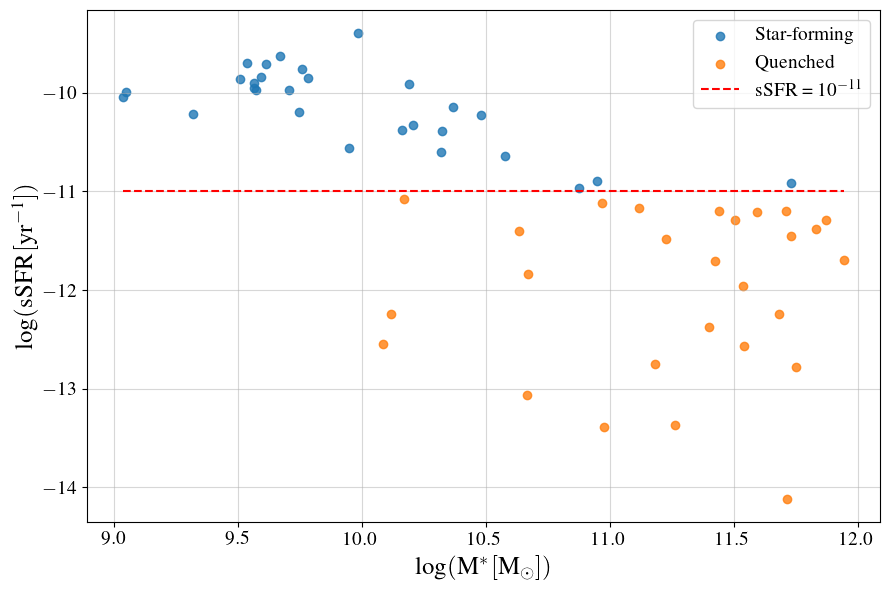}
    \includegraphics[clip=true, trim= 1mm 3mm 8mm 3mm, width=0.49\linewidth]{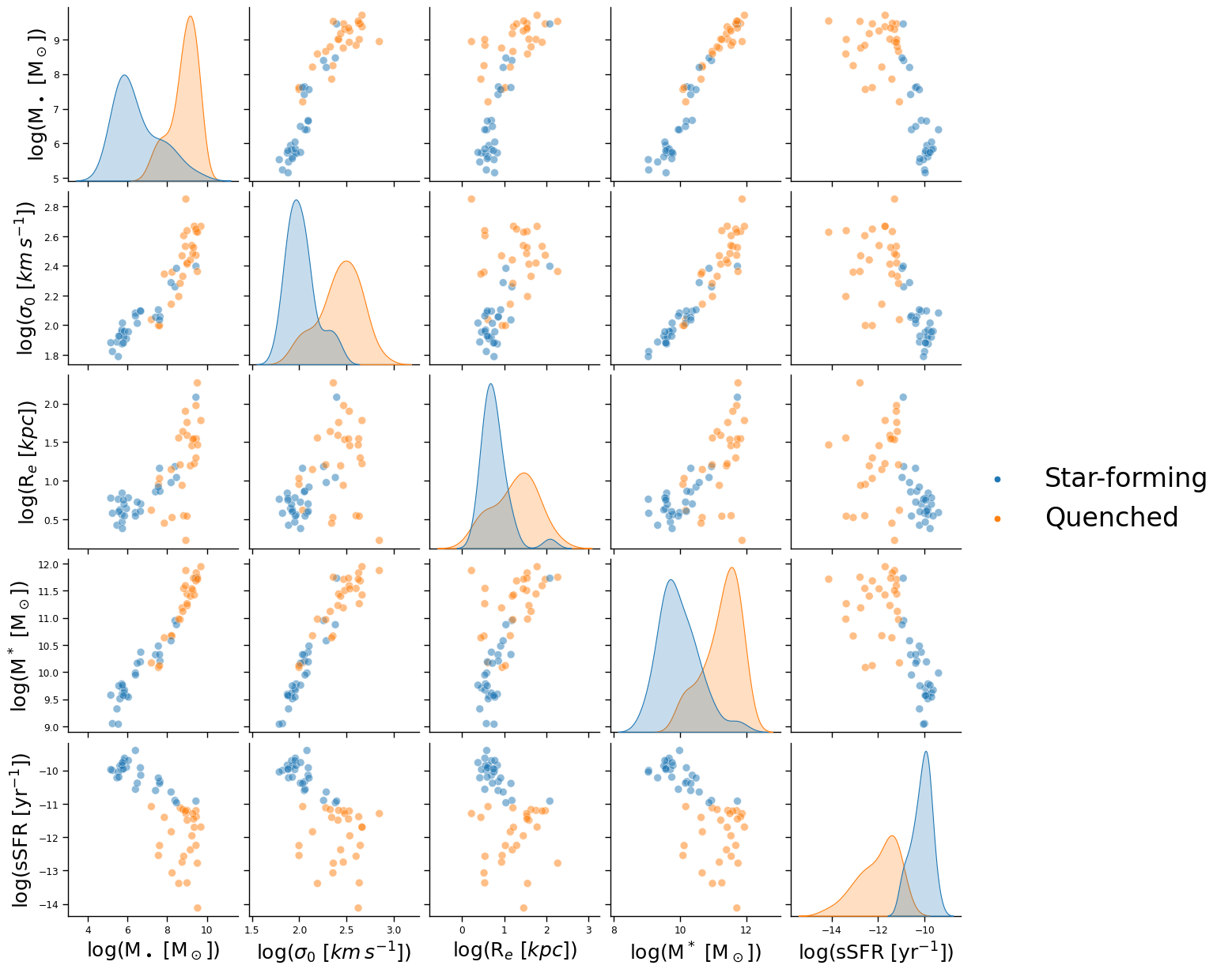}
    \caption{
    \emph{Left:} A plot of specific star-formation rate vs.\ stellar mass.
    Here, we use a cut (\textcolor{red}{{\hdashrule[0.35ex]{8mm}{1pt}{1pt}}}) at $\log(\textrm{sSFR}/\textrm{yr}^{-1})=-11$ to demarcate the 28 star-forming galaxies (\textcolor{matplotlib_blue}{$\bullet$}) from the 27 quenched galaxies (\textcolor{matplotlib_orange}{$\bullet$}).
    \emph{Right:} A pairplot of all the investigated parameters for the galaxies in this study.
    This pairplot illustrates the effectiveness of the $\log(\textrm{sSFR}/\textrm{yr}^{-1})=-11$ cut, which creates a clear bimodal distribution across all of the variables between star-forming galaxies and quenched galaxies.
    }
    \label{fig:categorize}
\end{figure*}

We begin our study by looking at 55 simulated galaxies that reached $z=0$ and grouping them into 28 star-forming galaxies and 27 quenched galaxies.
We classify the galaxies as either star-forming or quenched at $z=0$ by a cut in the SFR at $\log(\textrm{sSFR}/\textrm{yr}^{-1})=-11$ (see the \emph{left} plot in Figure~\ref{fig:categorize}).
This cut is such that it produces a clear bimodal distribution across all of the galaxy parameters (see the \emph{right} plot in Figure~\ref{fig:categorize}).
Coincidentally, this sSFR cut also divides our sample almost in half (28 vs.\ 27).
We emphasize that our goal is not to reproduce the exact observational distributions, but to assess whether simulated and observed systems in analogous evolutionary states exhibit consistent causal directions.

\begin{figure}
    \centering
    \includegraphics[clip=true, trim= 0mm 0mm 0mm 0mm, width=\linewidth]{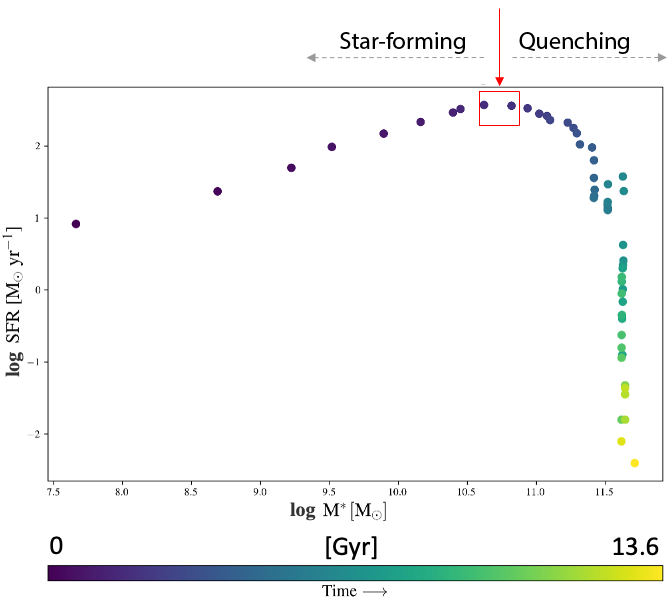}
    \caption{
    Evolution of a typical NIHAO galaxy that transitions from star-forming to quenching (one of 27).
    The plot shows the star-formation rate as a function of stellar mass.
    The stellar mass increases approximately monotonically as a function of time (indicated by the color of the marker).
    The star-formation rate initially increases until it reaches a maximum and then it steadily decreases.
    We use this turnover in the star-formation rate as a discriminator between a galaxy's star-forming epoch and its quenching epoch.
    }
    \label{fig:sfr_peak}
\end{figure}

Over the evolution of our simulated galaxies, we track the SFR of our galaxies as a function of time (Figure~\ref{fig:sfr_peak}).
At $z=0$, some galaxies have not yet reached their peak SFR; these galaxies are still in their ``star-forming'' phases.
Conversely, other galaxies at $z=0$ have already reached their peak SFR and are now on the decline (Figure~\ref{fig:sfr_peak}); these galaxies have reached their ``quenched'' phase.
This turnover in the star-formation rate helps us distinguish between a galaxy's period of active star formation and its subsequent period of quenching, where star formation slows or stops.

During the quenching phase of our simulated galaxies, we do witness brief periods of star-formation rejuvenation.
We have explicitly examined the prevalence and strength of star-formation rejuvenation events across the full simulated sample.
We define a rejuvenation episode as a secondary increase in SFR following the primary peak.
Such events are generally weak and short-lived: the median increase in SFR during a rejuvenation episode is only 0.14\,dex, and these fluctuations typically persist for a single simulation snapshot.
In no case does a rejuvenation episode approach the amplitude of the primary star-formation peak or introduce ambiguity in the identification of snapshot with the maximum SFR ($T_{\rm peak}$).
Consequently, while minor rejuvenation can temporarily modulate the inferred causal directions, these effects are transient and do not affect our classification of galaxies into star-forming and quenched phases or the overall temporal trends reported here.
Finally, our strategy of targeting the five snapshots before the global maxima of SFR (where star formation is most intense) and the very last snapshots leading up to $z=0$ (where we know the selected galaxy is quenched in the end) in our causal analyses is designed specifically for maximum contrast and have the highest chance of avoiding rejuvenation events.

The NIHAO galaxies analyzed here are drawn from cosmological zoom-in hydrodynamical simulations whose target halos were selected to span a wide range of halo masses, rather than to reproduce the stellar mass function of galaxies in a volume-complete sense.
As a result, the $z=0$ sample is intentionally non-uniform in $\log M^*$, and should not be interpreted as a representative census of galaxy populations.
Our analysis therefore conditions on this sample by construction.
Importantly, our principal results are driven by the \emph{temporal evolution within individual galaxies} and by \emph{contrasts between star-forming and quenched phases}, rather than by the relative number of systems at a given stellar mass.

Several recent cosmological simulations that explicitly resolve the multi-phase interstellar medium (ISM) and implement strong, localized stellar feedback predict highly bursty star formation histories in low-mass galaxies, with large fluctuations in star formation rates on $\sim$10--100\,Myr timescales \citep[e.g.,][]{El-Badry:2016,Ma:2018}.
These bursts arise from feedback-driven cycles of gas inflow, outflow, and re-accretion, and can significantly perturb the structural and kinematic properties of dwarf galaxies.
By contrast, the NIHAO simulations generally produce smoother star formation histories in this mass regime, reflecting differences in subgrid star formation and feedback prescriptions and the absence of an explicitly resolved cold ISM phase.

The impact of such burstiness on our results is expected to be limited, as our analysis emphasizes longer-timescale evolutionary trends and contrasts between star-forming and quenched phases, rather than short-timescale variability in the instantaneous star formation rate.
While bursty star formation may introduce additional scatter and temporal fluctuations in quantities such as $\sigma_0$ or SMBH accretion on Myr timescales, it is unlikely to qualitatively alter the inferred causal structure when averaged over the extended epochs considered here.
Nonetheless, we note that applying causal discovery methods to simulations with fully resolved ISM physics would be a valuable avenue for future work to assess the sensitivity of causal inferences to feedback-driven star formation variability.

\subsection{Confirmation of a Causal Reversal}

\begin{figure*}
\centering
    \includegraphics[clip=true, trim= 0mm 0mm 0mm 0mm, width=0.82\linewidth]{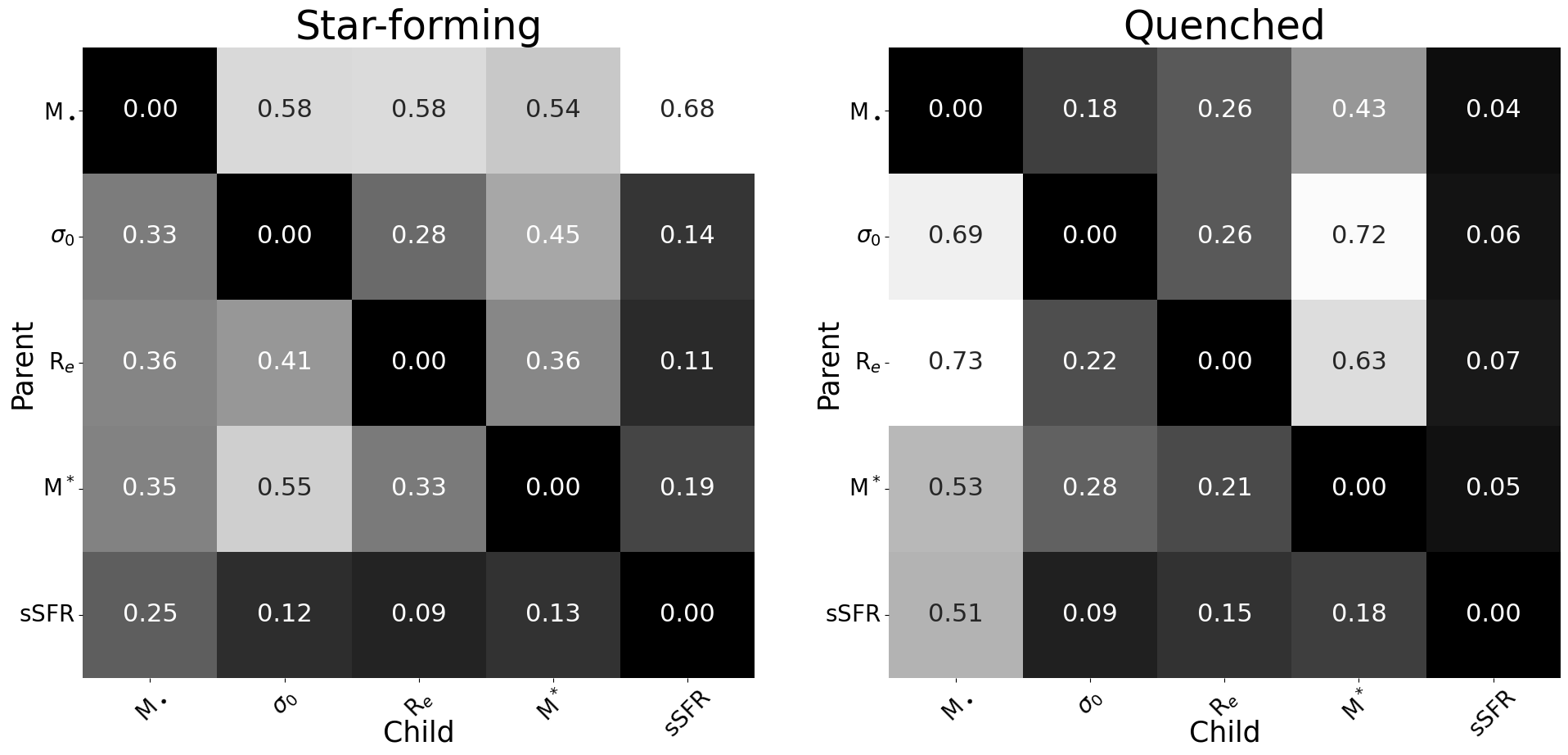}
    \includegraphics[clip=true, trim= 0mm 0mm 0mm 15mm, width=0.82\linewidth]{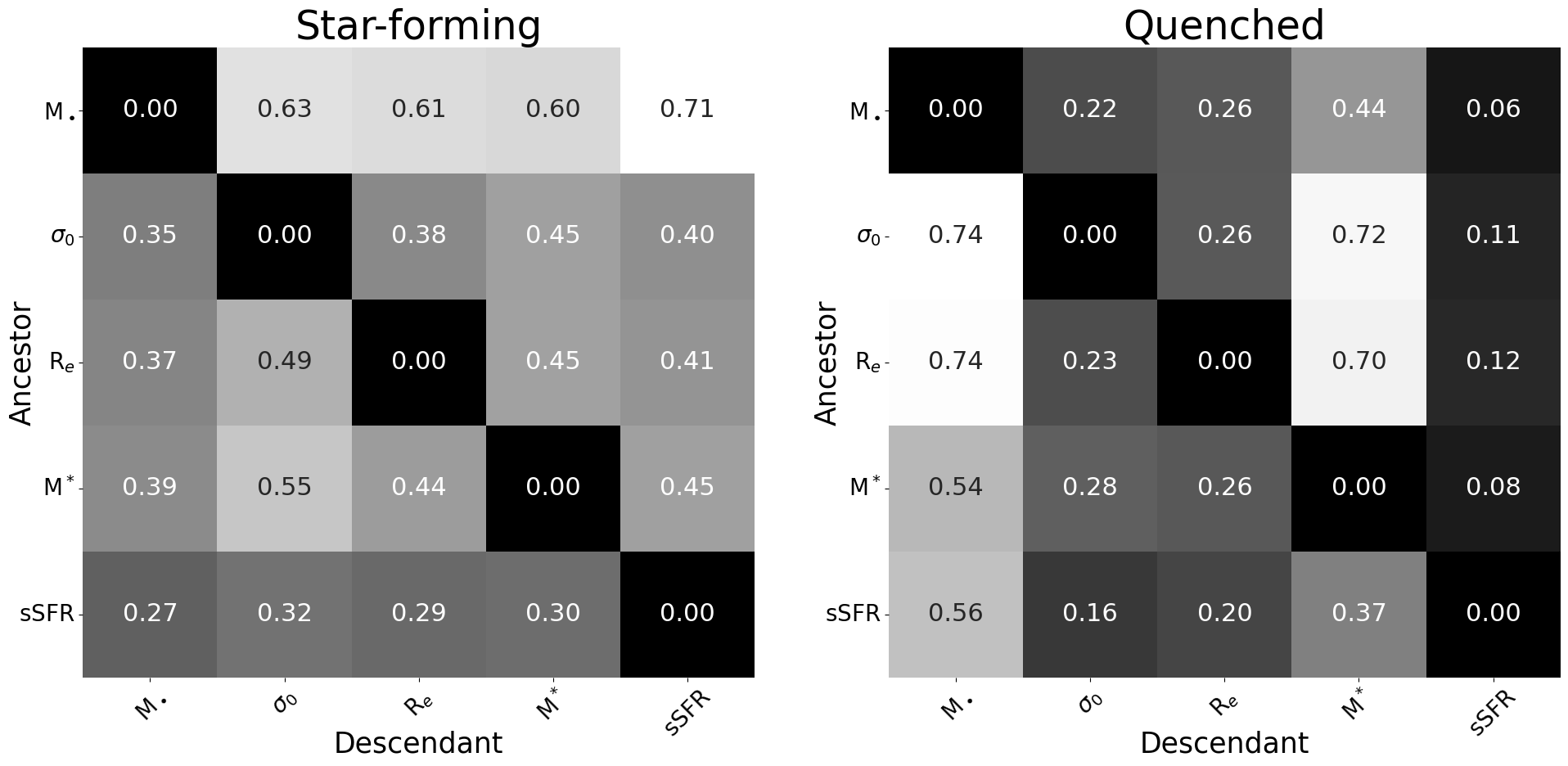}
    \includegraphics[clip=true, trim= 21.1cm 22mm 1mm 4mm, height=0.32\linewidth]{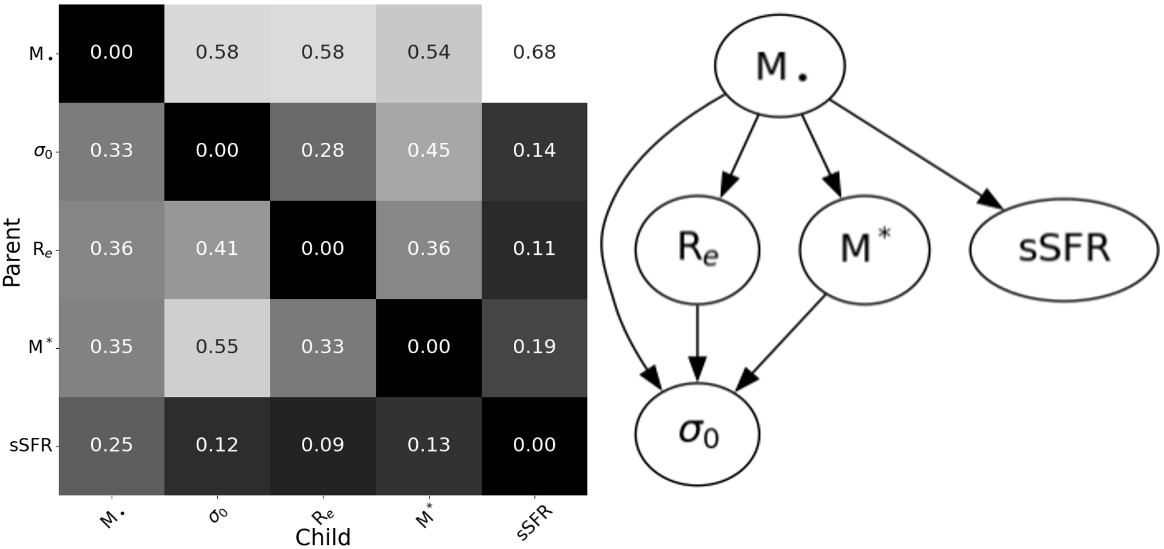}
    \includegraphics[clip=true, trim= 20.8cm 21mm 0mm 6mm, height=0.32\linewidth]{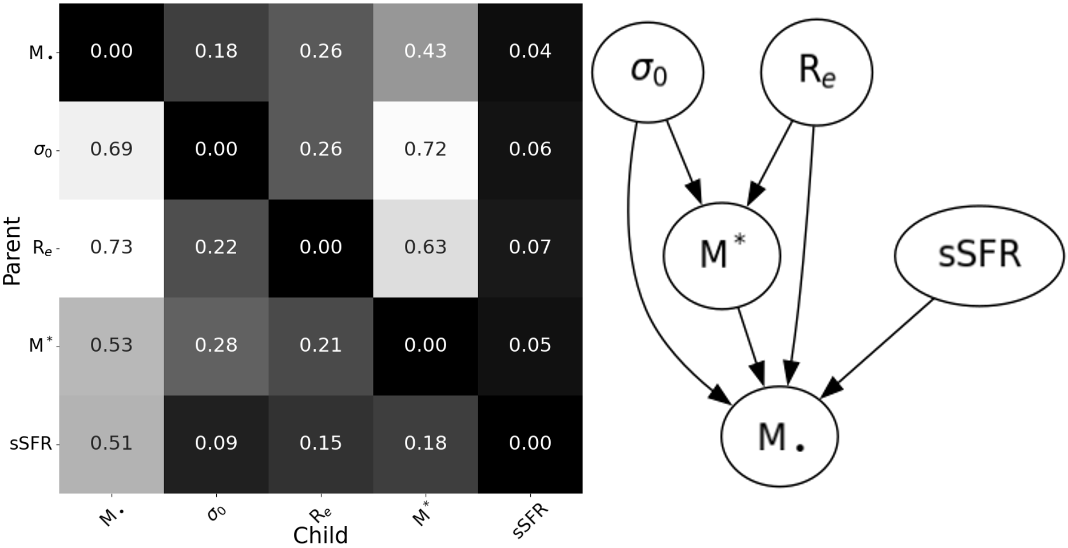}
    \caption{
    Edge (\emph{top row}) and path (\emph{middle row}) marginal matrices and the most probable directed acyclic graphs (DAGs; \emph{bottom row}) for the 28 simulated star-forming galaxies (\emph{left} column) and 27 simulated quenched galaxies (\emph{right} column) at $z=0$.
    These matrices clearly indicate a change in color between the first rows of each matrix and also between the first columns of each matrix.
    These changes represent higher probabilities for $M_\bullet$ \emph{causing} host galaxy properties for star-forming galaxies (\emph{left} matrices) and $M_\bullet$ being \emph{caused by} properties of their host galaxies in quenched galaxies (\emph{right} matrices).
    Indeed, the corresponding DAGs reflect $M_\bullet$ directly or indirectly \emph{causing} all the simulated properties of its host galaxy in star-forming galaxies (\emph{left} DAG) and $M_\bullet$ being \emph{caused by} all simulated properties of its host galaxy in quenched galaxies (\emph{right} DAG).
    }
    \label{fig:split}
\end{figure*}

Having grouped our simulated galaxies into star-forming and quenched, we then perform our causal discovery by computing the score-based exact posterior Bayesian distribution \citepalias[see][for complete details]{Jin:2025} for all 29,281 unique DAGs for our combination of five simulated parameters ($M_\bullet$, $\sigma_0$, $R_e$, $M^*$, and sSFR).
We present the sum contributions from all DAGs in edge and path marginal\footnote{The terminology \emph{edge} and \emph{path} marginals refer to the type of connections in a DAG. Edge marginals refer to all direct connections (i.e., an arrow directly from one node to another), while path marginals refer to all direct and indirect connections between nodes. For this reason, the variables in edge marginals are refereed to as ``parent/child'' versus ``ancestor/descendant'' in path marginals. Thus, path marginal values are always greater than or equal to edge marginal values because parents are always ancestors of their children, but ancestors are not always parents of their descendants.} matrices along with the corresponding single most probable DAG in Figure~\ref{fig:split}.
Focusing primarily on the parent/child relationships with $M_\bullet$, we find a clear trend of $M_\bullet$ being a \emph{parent} of all galaxy properties in star-forming galaxies, and oppositely a \emph{child} of all galaxy properties in quenched galaxies.\footnote{To clarify, we find that SMBHs are not significantly causing galaxy properties in quenched galaxies, \emph{as long as the galaxy remains quenched.} The caveat here is that if star formation gets reignited in a galaxy, this will allow the SMBH to reestablish causal influence on its host galaxy via AGN feedback. Indeed, \citet{Choi:2025} demonstrate that the shape of the stellar velocity dispersion functions of simulated pressure-supported galaxies is sensitive to the strength of AGN feedback.
However, as we mentioned in \S\ref{sec:SF}, our simulated galaxies typically undergo brief (one snapshot duration) and weak (0.14\,dex increase in SFR) rejuvenation events.}
These results are in tight agreement with the morphologically-separated (spiral galaxies vs.\ elliptical galaxies, respectively) samples in the observational study of \citetalias{Jin:2025}.

In Table~\ref{tab:compare}, we provide a comparison of all our edge and path marginals for simulated and the corresponding results for observed galaxies from \citetalias{Jin:2025}.
In each case, we group together and compare the values for \emph{spiral} galaxies from the observational study of \citetalias{Jin:2025} with \emph{star-forming} galaxies from our simulation study.
Similarly, we group together and compare the values for \emph{elliptical} galaxies from the observational study of \citetalias{Jin:2025} with \emph{quenched} galaxies from our simulation study.
Overall, we find close agreement in the causal directions and relative edge and path marginal values between the observed and simulated galaxies.

\begin{deluxetable*}{lccccc}
\tablecolumns{6}
\tablecaption{Comparison Between Observed and Simulated Galaxies at $z=0$}\label{tab:compare}
\tablehead{
\colhead{} & \multicolumn{5}{c}{Child} \\
\cline{2-6}
\colhead{Parent} & \colhead{$M_\bullet$} & \colhead{$\sigma_0$} & \colhead{$R_e$} & \colhead{$M^\ast$} & \colhead{sSFR}
}
\startdata
$M_\bullet$ & \nodata & $(0.62,0.58),(0.19,0.18)$ & $(0.51,0.58),(0.30,0.26)$ & $(0.69,0.54),(0.22,0.43)$ & $(0.35,0.68),(0.08,0.04)$ \\
$\sigma_0$ & $(0.22,0.33),(0.78,0.69)$ & \nodata & $(0.10,0.28),(0.59,0.26)$ & $(0.11,0.45),(0.52,0.72)$ & $(0.23,0.14),(0.07,0.06)$ \\
$R_e$ & $(0.25,0.36),(0.15,0.73)$ & $(0.10,0.41),(0.17,0.22)$ & \nodata & $(0.09,0.36),(0.46,0.63)$ & $(0.13,0.11),(0.05,0.07)$ \\
$M^\ast$ & $(0.30,0.35),(0.49,0.53)$ & $(0.18,0.55),(0.22,0.28)$ & $(0.17,0.33),(0.54,0.21)$ & \nodata & $(0.50,0.19),(0.05,0.05)$ \\
sSFR & $(0.13,0.25),(0.76,0.51)$ & $(0.17,0.12),(0.11,0.09)$ & $(0.06,0.09),(0.16,0.15)$ & $(0.20,0.13),(0.13,0.18)$ & \nodata \\
\hline
\colhead{} & \multicolumn{5}{c}{Descendant} \\
\cline{2-6}
\colhead{Ancestor} & \colhead{$M_\bullet$} & \colhead{$\sigma_0$} & \colhead{$R_e$} & \colhead{$M^\ast$} & \colhead{sSFR}\\
\hline
$M_\bullet$ & \nodata & $(0.71,0.63),(0.20,0.22)$ & $(0.63,0.61),(0.43,0.26)$ & $(0.69,0.60),(0.37,0.44)$ & $(0.74,0.71),(0.09,0.06)$ \\
$\sigma_0$ & $(0.25,0.35),(0.79,0.74)$ & \nodata & $(0.30,0.38),(0.76,0.26)$ & $(0.30,0.45),(0.72,0.72)$ & $(0.41,0.40),(0.16,0.11)$ \\
$R_e$ & $(0.30,0.37),(0.38,0.74)$ & $(0.36,0.49),(0.20,0.23)$ & \nodata & $(0.33,0.45),(0.46,0.70)$ & $(0.40,0.41),(0.09,0.12)$ \\
$M^\ast$ & $(0.31,0.39),(0.56,0.54)$ & $(0.31,0.55),(0.27,0.28)$ & $(0.37,0.44),(0.54,0.26)$ & \nodata & $(0.59,0.45),(0.11,0.08)$ \\
sSFR & $(0.22,0.27),(0.80,0.56)$ & $(0.37,0.32),(0.23,0.16)$ & $(0.32,0.29),(0.57,0.20)$ & $(0.35,0.30),(0.51,0.37)$ & \nodata \\
\enddata
\tablecomments{
Edge (\emph{top}) and path (\emph{bottom}) marginal values.
The nested ordered pairs indicate the following: (\emph{observed} spiral galaxies, \emph{simulated} star-forming galaxies), (\emph{observed} elliptical galaxies, \emph{simulated} quenched galaxies).
Consider the parent/child relationship above: $P(M_\bullet\rightarrow\sigma_0)=(0.62,0.58),(0.19,0.18)$.
The first nested ordered pair demonstrates the probability that $M_\bullet$ is a parent of $\sigma_0$ in \emph{observed} spiral galaxies is 62\% vs.\ 58\% for \emph{simulated} star-forming galaxies.
The second nested ordered pair demonstrates the probability that $M_\bullet$ is a parent of $\sigma_0$ in \emph{observed} elliptical galaxies is 19\% vs.\ 18\% for \emph{simulated} star-forming galaxies.
Therefore, these comparisons show a tight agreement between observed and simulated galaxies, in the case of $P(M_\bullet\rightarrow\sigma_0)$.
All simulated values are obtained from our Fig.~\ref{fig:split}, while all observed values are imported from \citetalias[][their Fig.~4]{Jin:2025}.
}
\end{deluxetable*}

In particular, we focus on the edge marginals concerning the $M_\bullet$--$\sigma_0$ relation, i.e., $P(M_\bullet\rightarrow\sigma_0)$ and $P(\sigma_0\rightarrow M_\bullet)$.
We find $P(M_\bullet\rightarrow\sigma_0) = 62\%$ for \emph{observed} spiral galaxies vs.\ 58\% for \emph{simulated} star-forming galaxies.
Similarly, we find $P(M_\bullet\rightarrow\sigma_0) = 19\%$ for \emph{observed} elliptical galaxies vs.\ 18\% for \emph{simulated} quenched galaxies.
For the opposite causal direction $P(\sigma_0\rightarrow M_\bullet)$, we find 22\% for \emph{observed} spiral galaxies vs.\ 33\% for \emph{simulated} star-forming galaxies.
Similarly, we find $P(\sigma_0\rightarrow M_\bullet) = 78\%$ for \emph{observed} elliptical galaxies vs.\ 69\% for \emph{simulated} quenched galaxies.\footnote{The opposing causal directions of the marginals do not necessarily add up to one, i.e., $P(X\rightarrow Y)+P(X\leftarrow Y)\leq1$. This is because there may be some probability that $X$ and $Y$ are disconnected. Counterintuitively, the edge marginals for the null case (i.e., the posterior from a uniform prior without any data) \emph{do not} imply equal probabilities (i.e., $1/3$) for $X\rightarrow Y$, $X\leftarrow Y$, and $X\indep Y$ ($X$ and $Y$ are independent). Although, $P(X\rightarrow Y)=P(X\leftarrow Y)$ must be satisfied in the null case. 
The \emph{edge} marginals for the null case  without any data for five variables is $P(X\rightarrow Y)=P(X\leftarrow Y)=30\%$ and $P(X\rightarrow Y)=P(X\leftarrow Y)=40\%$ for \emph{path} marginals.
}
These comparisons demonstrate a dominant preferential causal direction from $M_\bullet\rightarrow\sigma_0$ for both \emph{observed} spiral galaxies and \emph{simulated} star-forming galaxies.
Complementarily, we find a dominant preferential causal direction from $\sigma_0\rightarrow M_\bullet$ for both \emph{observed} elliptical galaxies and \emph{simulated} quenched galaxies.

\begin{figure}
\centering
\includegraphics[clip=true, trim=1mm 3mm 8mm 3mm, width=\linewidth]{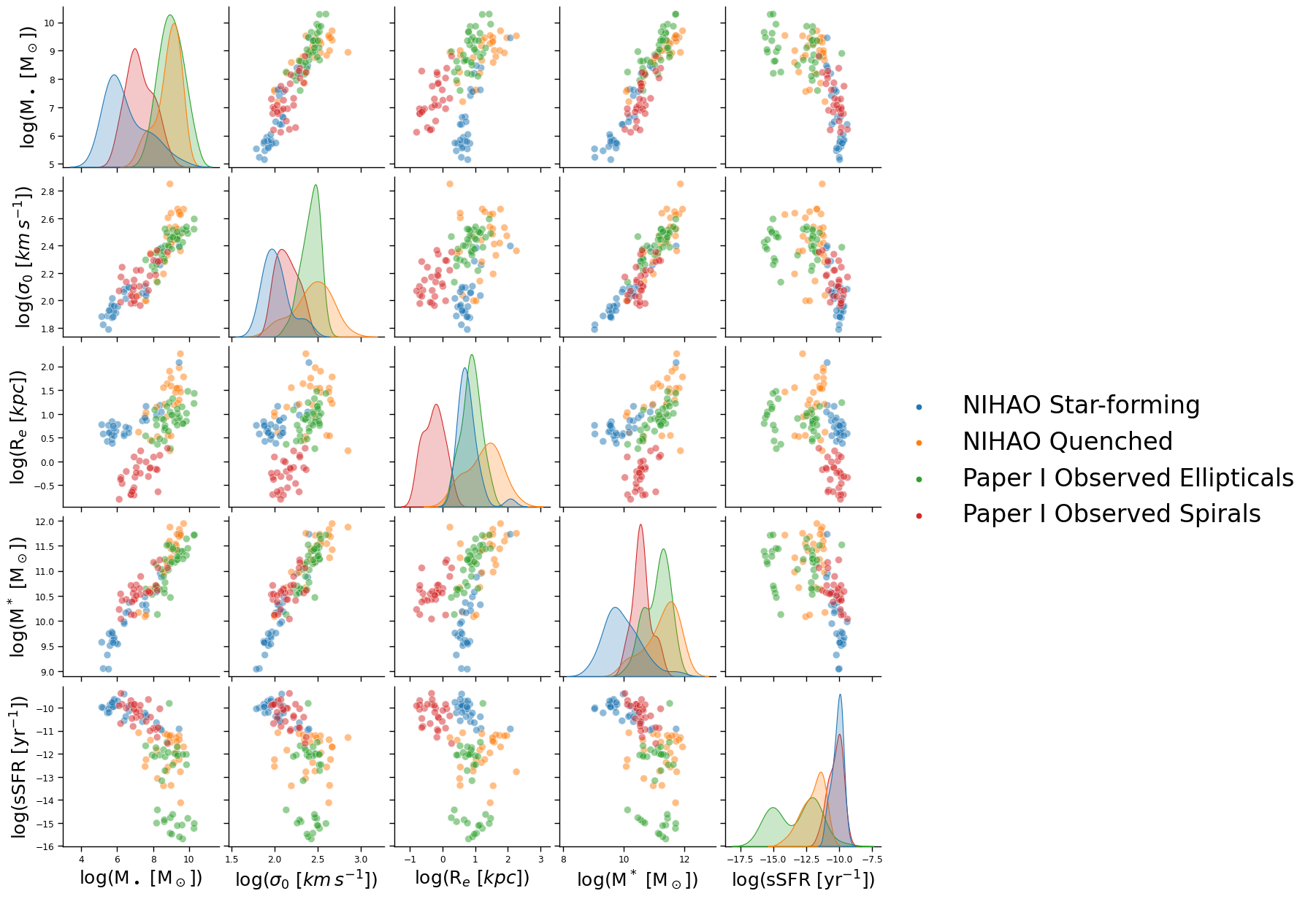}
\caption{
Comparison pairplot with the observational galaxy sample used in \citetalias{Jin:2025}, restricted to the same five variables considered in the present causal analysis ($M_\bullet$, $\sigma_0$, $R_e$, $M^\ast$, and sSFR).
The sample is separated into spiral galaxies (\textcolor{matplotlib_red}{$\bullet$}) and elliptical galaxies (\textcolor{matplotlib_green2}{$\bullet$}), with lenticulars excluded for clarity.
This figure data is analogous to Fig.~2 from \citetalias{Jin:2025}, but restricted to the parameter set used here.
Also, we plot again the star-forming galaxies (\textcolor{matplotlib_blue}{$\bullet$}) and quenched galaxies (\textcolor{matplotlib_orange}{$\bullet$}) from the pairplot in Fig.~\ref{fig:categorize} to show a visual comparison and facilitate interpretation of the posterior marginal comparisons presented in Table~\ref{tab:compare}.
}
    \label{fig:real}
\end{figure}

To aid interpretation of the comparison between simulated and observed marginal values in Table~\ref{tab:compare}, we include in Fig.~\ref{fig:real} a pairplot of the observational sample, restricted to the same five variables used in the causal analysis and separated by morphology (spirals versus ellipticals).
While the observational and simulated samples are not distribution-matched by construction, this visualization demonstrates that star-forming simulated galaxies and observed spirals, as well as quenched simulated galaxies and observed ellipticals, occupy broadly comparable regions of parameter space (compare with the pairplot from Fig.~\ref{fig:categorize}).
Differences in the detailed distributions are therefore expected and reflect the distinct selection functions of the simulations and observations, but do not preclude meaningful comparison of the inferred causal structures.

\subsection{Causal Reversal in the $M_\bullet$--$\sigma_0$ Relation}\label{sec:temp}

\begin{figure*}
\centering
\includegraphics[clip=true, trim=4mm 1mm 2mm 37mm, width=\linewidth]{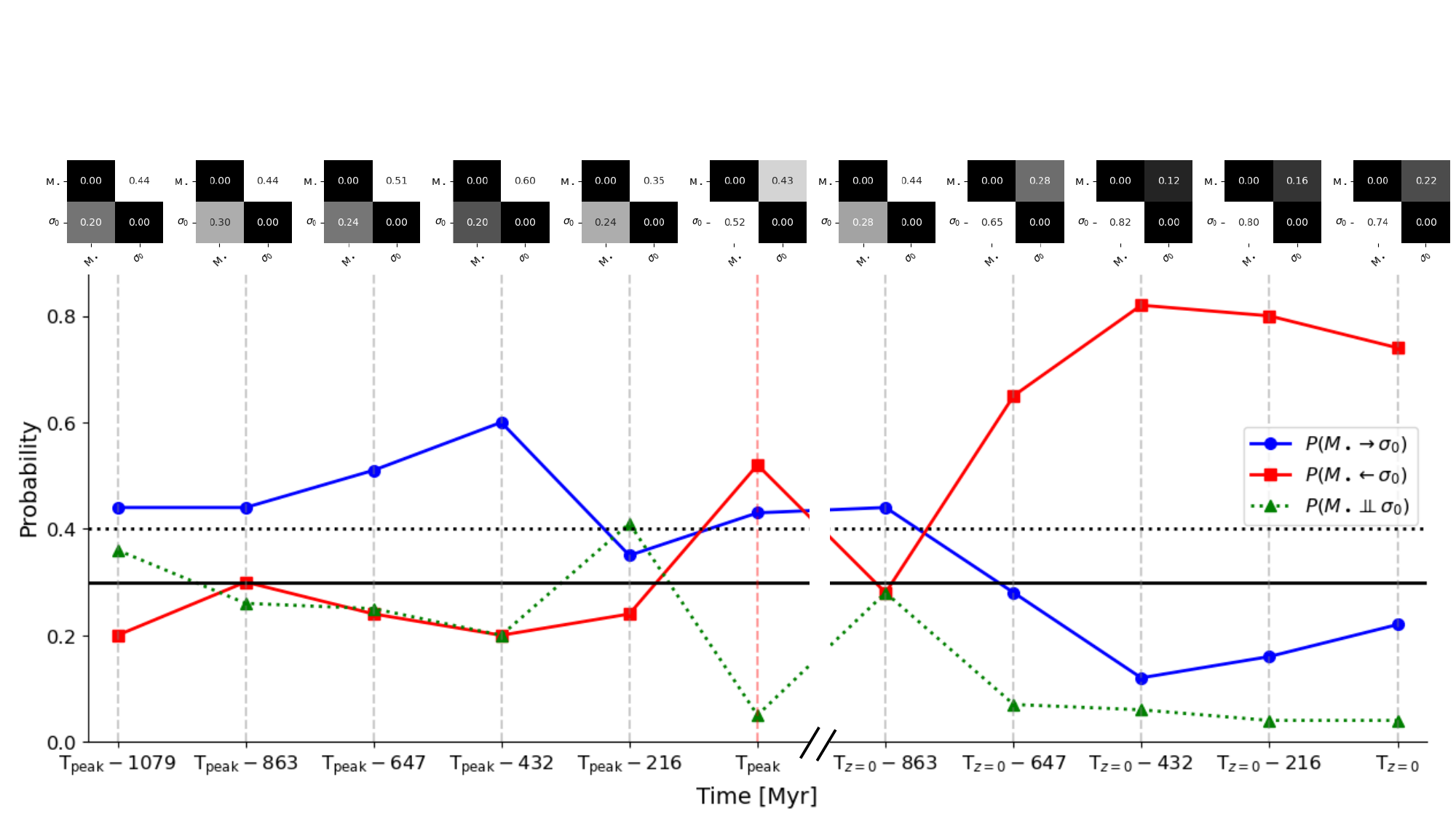}
\caption{Time evolution of the causal $M_\bullet$--$\sigma_0$ relation.
Here, we reduce the full $5\times5$ edge marginal matrices to only the $2\times2$ matrices concerning SMBH mass and central stellar velocity dispersion for our simulated galaxies, with 27 galaxies in our sample completing their evolution from star-forming to quenched by $z=0$.
These marginal matrices are listed across the top of the plot, with $x$-axes depicting causal children and the $y$-axes depicting causal parents.
We select eleven matrices from successive snapshots, centered at the peak of star formation at $T_\mathrm{peak}$ (the vertical \textcolor{lightcoral}{{\hdashrule[0.35ex]{8mm}{1pt}{1mm}}}).
The plot below the matrices connects the snapshot times of the matrices via the vertical \textcolor{lightgray}{{\hdashrule[0.35ex]{8mm}{1pt}{1mm}}}.
For each snapshot, we illustrate the probabilities of the $P(M_\bullet\rightarrow\sigma_0)$ causal direction (\textcolor{blue}{$\bullet$} connected by \textcolor{blue}{{\hdashrule[0.35ex]{8mm}{1pt}{}}}), its opposing $P(M_\bullet\leftarrow\sigma_0)$ represented by \textcolor{red}{$\blacksquare$} connected by \textcolor{red}{{\hdashrule[0.35ex]{8mm}{1pt}{}}}, and the probability, $P(M_\bullet\indep\sigma_0)$, that $M_\bullet$ and $\sigma_0$ are independent (\textcolor{matplotlib_green}{$\blacktriangle$} connected by \textcolor{matplotlib_green}{{\hdashrule[0.35ex]{8mm}{1pt}{1pt}}}); $P(M_\bullet\rightarrow\sigma_0)+P(M_\bullet\leftarrow\sigma_0)+P(M_\bullet\indep\sigma_0)=1$.
The horizontal lines at $P=0.4$ ({{\hdashrule[0.35ex]{8mm}{1pt}{1pt}}}) and $P=0.3$ ({{\hdashrule[0.35ex]{8mm}{1pt}{}}}) represent the null probabilities for the independency case and causal directional cases, respectively.
Thus, significance occurs when the plotted solid lines are further away from the solid horizontal line and the plotted dotted line is significantly different from the horizontal dotted line.
For example, the snapshot at $T_\mathrm{peak}-216$\,Myr demonstrates a period of transition that lacks any meaningful causal information because all values are near their null values.
}
    \label{fig:time}
\end{figure*}

In Figure~\ref{fig:time}, we present a graphical representation of the time evolution of the causal $M_\bullet$--$\sigma_0$ relation, relative to a galaxy's age at the peak of star-formation ($T_\mathrm{peak}$) and at $z=0$ ($T_{z=0}$).
This timeline of causality demonstrates how $P(M_\bullet\rightarrow\sigma_0)>P(M_\bullet\leftarrow\sigma_0)$ is consistent from snapshots $T_\mathrm{peak}-1079$\,Myr to $T_\mathrm{peak}-$432\,Myr.
The snapshots at $T_\mathrm{peak}$, and closest to the peak of star formation (from $T_\mathrm{peak}-216$\,Myr to $T_{z=0}-863$\,Myr), show a period of transition where $P(M_\bullet\rightarrow\sigma_0)>P(M_\bullet\leftarrow\sigma_0)$ switches to $P(M_\bullet\rightarrow\sigma_0)<P(M_\bullet\leftarrow\sigma_0)$, and back again.
This period of transition could be analogous to the variability in a galaxy as its black hole grows and it moves toward quiescence, as exhibited by its fluctuating fraction of cool gas \citep{Wang:2024b}.
Following the transition, the snapshots settle to $P(M_\bullet\rightarrow\sigma_0)<P(M_\bullet\leftarrow\sigma_0)$ from snapshots $T_{z=0}-647$\,Myr to $T_{z=0}$.
Therefore, we consider our simulated galaxies to firmly have the $M_\bullet\rightarrow\sigma_0$ causal direction during their evolution \emph{up to} 432\,Myr before their peak star-formation rate, and decidedly flip to $M_\bullet\leftarrow\sigma_0$ for their futures \emph{beyond} 647\,Myr before $T_{z=0}$.
This temporal evolution of the $M_\bullet$--$\sigma_0$ relation provides a more finely resolved nature of the causal reversal discovered by \citetalias{Jin:2025} from analyzing observations of real galaxies segregated by morphology.

\section{Causally-informed Scaling Relations}\label{sec:scaling}

We use the linear regression software \texttt{Hyper-Fit} \citep{Robotham:2015,Robotham:2016} to accurately fit our relations from the observational data of \citetalias{Jin:2025} while fully considering uncertainties in all variables and intrinsic scatter.
Since we will be working entirely in logarithmic space\footnote{
We work in logarithmic space because the relevant physical relations are approximately power laws in linear space and the variables span several orders of magnitude, making their dependencies closer to linear when expressed in log form.
The linear--Gaussian assumption in our Bayesian network applies to the \emph{conditional} relationships between variables, not to their marginal distributions; thus, we do not require individual variables to be strictly Gaussian.
We do not Gaussianize or otherwise standardize the variables beyond taking logarithms, and non-Gaussian features (e.g., skewness or long tails) are absorbed into the intrinsic scatter of the conditional likelihoods rather than biasing the inferred causal directions.}, we make the following notations for simplicity: $\mathcal{M}_\bullet\equiv\log(M_\bullet/\textrm{M}_\sun)$ and $\mathcal{S}_0\equiv\log(\sigma_0/\textrm{km\,s}^{-1})$.
We will consider $\mathcal{M}_\bullet$ to be the black hole mass derived from the causal $M_\bullet\leftarrow\sigma_0$ direction.
Conversely, we will represent the anti-causal ($M_\bullet\rightarrow\sigma_0$) as $\mathcal{M}_\bullet^\prime$.
We will denote the intrinsic scatter on the dependent variable in all relations as $\epsilon$, which \emph{is already included} in our relations.
We note that the $M_\bullet$--$M^\ast$ relation also demonstrates an equivalent causal reversal (see Fig.~\ref{fig:split}).
Because the $M_\bullet$--$M^\ast$ relation is also useful for predicting black hole masses at high-$z$, we will also similarly present its relations here with $\mathcal{M}^\ast\equiv\log(M^\ast/\textrm{M}_\sun)$.

\subsection{Baseline: All Galaxies}

Most $M_\bullet$--$\sigma_0$ relations in the literature are fit to varied samples of all galaxy types grouped together in heterogeneous mixes of morphological types.
However, as we have demonstrated with our causal analysis, this is not appropriate due to the mixing of the causal and anti-causal directions from quenched and star-forming galaxies, respectively.
For comparison purposes, we fit the ``wrong'' ($\mathcal{M}_\bullet^\prime$) black hole mass as a function of central stellar velocity dispersion from the sample of 101 galaxies (spirals, lenticulars, and ellipticals) from \citetalias{Jin:2025}.
\begin{align}
\mathcal{M}_\bullet^\prime&=(8.25\pm0.46)+(4.92\pm0.02)(\mathcal{S}_0-2.28\label{eqn:sigma-wrong})\\
\epsilon&=0.46\pm0.04\nonumber
\end{align}
Similarly, we also fit the wrong black hole mass as a function of stellar mass.
\begin{align}
\mathcal{M}_\bullet^\prime&=(8.01\pm0.66)+(1.52\pm0.01)(\mathcal{M}^\ast-10.63)\label{eqn:stellar-wrong}\\
\epsilon&=0.66\pm0.05\nonumber
\end{align}

\subsection{Spiral (i.e., Star-forming) Galaxies}

For the ``correct'' causal direction for star-forming galaxies, we consider the sample of 28 spiral galaxies from \citetalias{Jin:2025} and fit $\sigma_0$ as a function of $M_\bullet$, and then invert the relation.
\begin{align}
\mathcal{S}_0&=(2.12\pm0.08)+(0.12\pm0.01)(\mathcal{M}_\bullet-7.07)\nonumber\\
\epsilon&=0.08\pm0.02\nonumber\\
\mathcal{M}_\bullet&=7.07+\frac{\mathcal{S}_0-(2.12\pm0.08)}{0.12\pm0.01}
\label{eqn:spiral-inverted}
\end{align}
A similar fit and subsequent inversion is performed for stellar mass.
\begin{align}
\mathcal{M}^\ast&=(10.53\pm0.19)+(0.31\pm0.01)(\mathcal{M}_\bullet-7.07)\nonumber\\
\epsilon&=0.19\pm0.03\nonumber\\
\mathcal{M}_\bullet&=7.07+\frac{\mathcal{M}^\ast-(10.53\pm0.19)}{0.31\pm0.01}
\label{eqn:stellar-inverted}
\end{align}

\subsection{Elliptical (i.e., Quenched) Galaxies}

For quenched galaxies with the correct causal direction from $\sigma_0\rightarrow M_\bullet$, we directly fit a relation for the $M_\bullet$--$\sigma_0$ relation from the 35 elliptical galaxies in \citetalias{Jin:2025}.
\begin{align}
\mathcal{M}_\bullet&=(9.17\pm0.39)+(5.13\pm0.01)(\mathcal{S}_0-2.44)\label{eqn:elliptical-sigma}\\
\epsilon&=0.38\pm0.05\nonumber
\end{align}
Similarly, we can directly fit $M_\bullet$ as a function of $M^\ast$.
\begin{align}
\mathcal{M}_\bullet&=(9.17\pm0.38)+(1.37\pm0.00)(\mathcal{M}^\ast-11.22)\\
\epsilon&=0.38\pm0.06\nonumber
\end{align}

\subsection{A Correction for Spiral/Star-forming Galaxies}

If a black hole mass were improperly fit based on Equation~\ref{eqn:sigma-wrong}, we can combine it with Equation~\ref{eqn:spiral-inverted} to create the following equation to causally correct the black hole mass.
\begin{align}
\mathcal{S}_0&=A+B(\mathcal{M}_\bullet-7.07\nonumber)\\
\mathcal{M}_\bullet^\prime&=\alpha+\beta(\mathcal{S}_0-2.28)\nonumber\\
\mathcal{M}_\bullet&=7.07+\frac{\mathcal{M}_\bullet^\prime-\alpha-\beta(A-2.28)}{\beta B}\nonumber\\
&=7.07+\frac{\mathcal{M}_\bullet^\prime-(7.46\pm0.61)}{0.59\pm0.04}
\label{eqn:sigma-correct}
\end{align}
Here, there is a unique value where $\mathcal{M}_\bullet^\prime=\mathcal{M}_\bullet$ at $\approx$8.02, below which black hole masses are over-predicted and above which black hole masses are under-predicted.
Likewise, we can do the same for incorrect masses from Equation~\ref{eqn:stellar-wrong} by combining with Equation~\ref{eqn:stellar-inverted} to create the following correcting formula for black hole masses derived from stellar masses.
\begin{align}
\mathcal{M}^\ast&=A+B(\mathcal{M}_\bullet-7.07\nonumber)\\
\mathcal{M}_\bullet^\prime&=\alpha+\beta(\mathcal{M}^\ast-10.63)\nonumber\\
\mathcal{M}_\bullet&=7.07+\frac{\mathcal{M}_\bullet^\prime-\alpha-\beta(A-10.63)}{\beta B}\nonumber\\
&=7.07+\frac{\mathcal{M}_\bullet^\prime-(7.86\pm0.72)}{0.47\pm0.01}
\end{align}
Here, there is a unique value where $\mathcal{M}_\bullet^\prime=\mathcal{M}_\bullet$ at $\approx$8.56, below which black hole masses are over-predicted and above which black hole masses are under-predicted.

\section{Discussion}\label{sec:discussion}

\subsection{Predicting SMBH Masses}

A causation always implies a correlation\footnote{In principle, causation implies a correlation, however, the correlation may not always be detectable. Non-linear relations, countervailing causes, and dichotomous or manipulated samples may make it difficult to determine a significant correlation (specifically, linear correlation).}, but a correlation does not always imply a causation.
For the $M_\bullet$--$\sigma_0$ relation, the results of \citetalias{Jin:2025} and this study indicate that $M_\bullet$ is \emph{always correlated} with $\sigma_0$, but only in the case of quenched galaxies (i.e., elliptical galaxies) is $\sigma_0$ \emph{predominantly a cause} of $M_\bullet$.
This result is also suggested from the pure study of correlations via linear regressions.
\citet{Sahu:2019b} found that the intrinsic scatter ($\epsilon$) of the $M_\bullet$--$\sigma_0$ relation is significantly higher for spiral galaxies ($\epsilon=0.67$\,dex) than for elliptical galaxies ($\epsilon=0.31$\,dex).
Furthermore, the existence of morphologically-aware black hole mass scaling relations also supports the existence of distinct evolutionary pathways for late- vs.\ early-type galaxies \citep{Graham:2013,Scott:2013,Savorgnan:2013,Savorgnan:2016,Davis:2018,Davis:2019,Sahu:2019,Sahu:2019b,Sahu:2020,Sahu:2022,Sahu:2022b,Graham:2023}.
In a more absolute scenario, \citet{Chen:2025} present evidence for naked ``little red dots'' (LRDs) with essentially no detectable host galaxy \citep[cf.][]{Zhang:2025}.
Thus, black holes would be the answer to the \emph{chicken-or-the-egg dilemma} of which came first, making black holes the default cause of their subsequent host galaxies (in young star-forming galaxies).

\citet{Novak:2006} explored several black hole mass scaling relations (including the $M_\bullet$--$\sigma_0$ relation) in search of the ``true'' (i.e., causal) correlations of SMBHs.
They emphasized that in order to arrive at the correct theory governing black hole mass scaling relations, it is essential to determine the correlation with the smallest intrinsic residual variance, which will have the best chance of being causally related to SMBH mass.
Furthermore, \citet{Novak:2006} discussed that knowledge of the direction of the causal link between SMBH mass and galactic properties is required to correctly sort variables into their roles as ``dependent'' or ``independent.''
In the absence of such knowledge, it is necessary to always treat variables $x$ and $y$ symmetrically in linear regressions.
Lastly, \citet{Novak:2006} describe the different goals of a theorist vs. an observer; a theorist is always concerned with obtaining the causally-motivated relation (i.e., the lowest intrinsic scatter), while the observer is only interested in predicting black hole masses (i.e., $M_\bullet$ is always the dependent variable).

From our results, we find that the ``theorist'' and the ``observer'' will only agree on applying the $M_\bullet$--$\sigma_0$ relation to elliptical galaxies.
For elliptical galaxies, the intrinsic scatter will be minimized and both the theorist and the observer will use $\sigma_0$ to predict $M_\bullet$.
For spiral galaxies, the theorist will prefer to use $\sigma_0$ as the dependent variable with $M_\bullet$ as the independent variable, and the observer will probably find some other relation that minimizes the predicted error on $M_\bullet$.
For the observer, obtaining more accurate $M_\bullet$ estimates can be more readily obtained by switching or adding more variables \citep{Jin:2023,Davis:2023,Davis:2024}.
Overall, our method of applying causal discovery provides a more definitive method of determining causal correlations beyond simply performing linear regressions and searching for the lowest intrinsic scatter.

Recent observational and theoretical studies have suggested that the mass \citep[and/or concentration;][]{Nino:2025} of the host dark matter halo may also correlate with, or even predict, SMBH mass \citep[e.g.,][]{Voit:2024,Zhang:2024}.
Whether this relation reflects a direct causal influence of halo-scale structure on SMBH growth, or instead arises indirectly through baryonic processes that link halo mass to galaxy-scale properties (such as stellar mass, gas supply, or central velocity dispersion), remains an open question.
Within a causal framework, halo mass could plausibly act as an upstream variable that shapes the long-term gas accretion history and merger environment of the galaxy, thereby influencing SMBH growth without requiring a direct physical coupling across vastly different spatial scales.
Distinguishing between direct halo $\rightarrow$ SMBH causation and mediated pathways (halo $\rightarrow$ galaxy $\rightarrow$ SMBH) represents a promising avenue for future work, particularly with simulations that explicitly track halo assembly histories alongside galaxy and SMBH evolution.

\subsection{Implications for High-$z$ Galaxies}

Our results concerning the causal directions in the $M_\bullet$--$\sigma_0$ relation portend the need for careful consideration when predicting black hole masses at high redshifts.
The peak of star formation (also known as ``cosmic noon'') occurred at $z\sim2$ \citep{Madau:2014}.
With the \textit{James Webb Space Telescope (JWST)} now regularly measuring the mass of SMBHs at redshifts of at least $z\approx7$ \citep[e.g.,][]{Ubler:2023,Stone:2024,Inayoshi:2025}, it is crucial that the methods used to measure these SMBH masses are correctly applied.
Because these \textit{JWST}-discovered SMBHs all existed well before cosmic noon (more like cosmic morning), they come from an era of the Universe that was predominantly star-forming.

This star-forming dependence implies that na\"{i}vely applying the $M_\bullet$--$\sigma_0$ relation, and indirect methods that are calibrated to the $M_\bullet$--$\sigma_0$ relation like single-epoch spectra or reverberation mapping, to all galaxies is not causally motivated.
Based on our causal analysis shown in Fig.~\ref{fig:time} and described in \S\ref{sec:temp}, we find the causal reversal of the $M_\bullet$--$\sigma_0$ relation occurs around the peak of star formation, with transitions beginning 432\,Myr before $T_\mathrm{peak}$ and ending 647\,Myr before $T_{z=0}$.
Adapting this trend to the general population of galaxies, we estimate that this translates to $M_\bullet\rightarrow\sigma_0$ at $z\gtrsim2.1$ ($\lesssim$3.1\,Gyr \emph{after} the Big Bang) and $M_\bullet\leftarrow\sigma_0$ at $z\lesssim0.047$ ($\lesssim$0.65\,Gyr \emph{before} $T_{z=0}$).\footnote{Here, we specifically adopt the \hspace{-1.5mm}\citet{Planck:2020} cosmological parameters and specify cosmic noon at $z\approx1.9$ or $\approx$3.5\,Gyr after the Big Bang \citep{Madau:2014}.}
Thus, we find that use of the unmodified $M_\bullet$--$\sigma_0$ relation (and its derivatives) to predict black hole masses is causally-justified broadly at $z\lesssim0.047$ (luminosity distances closer than $\approx$210\,Myr).
The intermediate period at $0.047\lesssim z\lesssim2.1$ may allow for partial justification of causally-based $M_\bullet$ prediction, whereas epochs at $z\gtrsim2.1$ are likely not causally-supported.
Nonetheless, elliptical galaxies (or any types of quenched galaxies) at any redshift remain causally-motivated targets for black hole mass estimation.
This long transitionary period ($0.047\lesssim z\lesssim2.1$) lasts $\approx$10\,Gyr, which is consistent with the duration of galaxies in the slow-quenching tail (i.e., taking longer than 1\,Gyr to quench) that occupy a dominant fraction of the distribution \citep{Walters:2022}.

We have demonstrated from our simulations that star-forming galaxies at high redshifts host SMBHs that are not causally beholden to stellar velocity dispersion.
Therefore, we expect that there will be increased uncertainty in any SMBH mass derived from an uncorrected $M_\bullet$--$\sigma_0$ relation at high redshifts.
This increased uncertainty is due to incorrectly assuming that $\sigma_0$ causes $M_\bullet$ in star-forming galaxies.
We expect that this excess uncertainty may be contributing to the frequent identification of so-called ``overmassive'' black holes at high redshifts.
Notably, \citet{Bosman:2025} show that there does not appear to be any intrinsic differences between low-$z$ ($z<3$) and high-$z$ ($z>7$) quasars, suggesting that the inferred black hole masses are possibly incorrect at both epochs.

The existence of overmassive black holes at such early epochs of the Universe has sparked intense debates concerning early black hole seeding mechanisms \citep{Volonteri:2023}, LRDs \citep{Ananna:2024,Khan:2025}, obscuring interstellar media \citep{Gilli:2022}, and super-Eddington accretion \citep{Jeon:2023,Trinca:2024,Suh:2025}.
Instead, we offer the simple explanation that perhaps these claims of overmassive black holes should be taken with a grain of salt due to \emph{$M_\bullet$ being causally antecedent to $\sigma_0$, not a consequence of it in star-forming galaxies}.
Moreover, there are other reasons to question the existence of overmassive black holes in LRDs at high-$z$ \citep[e.g.,][]{Sacchi:2025,Wang:2025,Burke:2025,Diego:2025}.
Indeed, \citet{Gravity:2025} directly measured the broad-line region (BLR) dynamics of a quasar at $z=4$; they found its derived mass ``is an order of magnitude lower than that inferred from various single-epoch scaling relations, and implies that that the accretion is highly super-Eddington.''

Recent \textit{JWST} studies that account for selection effects find that the intrinsic SMBH--stellar mass relation at $z \gtrsim 6$ need not be strongly elevated, but instead exhibits substantially larger scatter relative to the local relation \citep{Silverman:2025}.
Similarly, \citet{Parlanti:2025} indicate that the masses of highly-accreting black holes at high redshifts, derived from single-epoch, may be systemically overestimated by an order of magnitude.
Additionally, \citet[][\S4.1, and references therein]{Hannah:2025} provide numerous explanations for why the masses of high-$z$ are likely overestimated.
Increasing doubt about the provenance of SMBHs in the early Universe will relieve the tension and need for exotic theories to explain their seemingly premature origins.

An important caveat concerns the interpretation of LRDs, a population of extremely compact, high-redshift sources with unusually broad Balmer emission lines, and seemingly unphysical ages \citep{Corredoira:2026}.
While some recent studies have suggested that LRDs may host actively accreting SMBHs with little or no detectable stellar component \citep[e.g.,][]{Juod:2025}, the nature of their host galaxies remains uncertain.
In particular, recent stacking analyses indicate that, if present, the stellar hosts of LRDs are extremely compact ($\sim$200\,pc) and lie closer to the size--mass relation of quiescent or dynamically hot systems than to that of typical star-forming galaxies \citep[e.g.,][]{Chen:2024,Zhang:2025}.
This raises the possibility that LRDs represent quenched galaxies, compact spheroids, or systems undergoing rejuvenation, rather than ordinary star-forming disks.
Our causally-informed correction to black hole mass estimates is most directly applicable to star-forming galaxies with extended stellar components, for which the causal direction $M_\bullet \rightarrow \sigma_0$ is supported by both observations and simulations.
If LRDs are instead hosted by quenched or dynamically hot systems, or if their stellar components are fundamentally distinct from those of typical star-forming galaxies, then the causal framework developed here may not apply.
We therefore regard LRDs as a special and potentially transitional population \citep{Hviding:2026} whose causal SMBH--galaxy connections remain to be established, and we emphasize that dedicated studies---combining resolved kinematics, star-formation diagnostics, and causal analysis---may be required to determine the appropriate scaling relations for these objects.

Moreover, recent JWST spectroscopic studies have begun to shed new light on the physical nature of LRDs.
In particular, high-quality spectra indicate that the broad-line wings in many LRDs are dominated by electron scattering in a dense, ionized medium rather than by bulk Doppler motions of virialized gas, implying intrinsic line widths far smaller than the apparent wings and correspondingly lower black hole masses \citep{Rusakov:2025}.
Such electron scattering signatures suggest that the line shapes reflect radiative transfer through nearly Compton-thick cocoons rather than the gravitational potential of an extended stellar system, and thus traditional virial mass estimators calibrated on galaxy nuclei may be inappropriate for these objects.
If LRDs indeed comprise young, rapidly growing SMBHs enshrouded in dense gas, their causal connection to stellar mass, central velocity dispersion, or the scaling relations derived for extended galaxies remains unclear.

\subsection{Redetermining the Masses of High-$z$ Black Holes}

For a demonstration of our causal correction to high-$z$ black hole masses, we apply our results to the sample of 13 black holes at $4.1<z<10.6$ from \citet{Maiolino:2024}.
\citet{Maiolino:2024} presents a collection of star-forming, late-type systems\footnote{Additionally, \citet{Ono:2025} find that the morphological properties of high-$z$ galaxies are statistically similar to local spiral galaxies.} with black holes that are overmassive with respect to local black hole to galaxy stellar mass ratios.
Their black hole masses are estimated via local single-epoch virial black hole mass relations \citep{Reines:2013,Reines:2015}.
However, these estimates are not directly derived from dynamical black hole masses.
Instead, reverberation-mapped black hole masses are calibrated to the local $M_\bullet$--$\sigma_0$ relation, then single-epoch spectroscopic black masses are correlated with reverberation mapping via the BLR radius--luminosity relation.
Therefore, single-epoch spectroscopic black hole masses are indirectly calibrated to the local $M_\bullet$--$\sigma_0$ relation.\footnote{In this way, it is perhaps helpful to think about the connection between 1.) dynamical black hole measurements, 2.) reverberation-mapped black hole masses, and 3.) single-epoch spectroscopic black hole masses as being analogous to successive rungs on the cosmic distance ladder. If there are inaccuracies in the lowest rung of the ladder, these errors will propagate to the higher rungs of the ladder.}
Furthermore, since the local $M_\bullet$--$\sigma_0$ relation is almost exclusively fit to a sample of star-forming plus quenched galaxies, this implies that our causal correction formalism can be used to correct for the magnitude of offset in the predicted black hole masses.

Although \citet{Maiolino:2024} do not have direct measurements of the central \emph{stellar} velocity dispersions for their galaxies, they are able to accurately measure the velocity dispersion of \emph{gas} from their high-resolution spectra.
From there, they are able to make small corrections \citep{Bezanson:2018} to closely approximate the stellar velocity dispersion.
The result is an effective central stellar velocity dispersion that can be used as a good proxy for the type of $\sigma_0$ that is used in local scaling relations.
We use these values with our causal $M_\bullet$--$\sigma_0$ relations to estimate revised black hole masses.

\begin{figure}
    \includegraphics[clip=true, trim= 3mm 3mm 3mm 3mm, width=\linewidth]{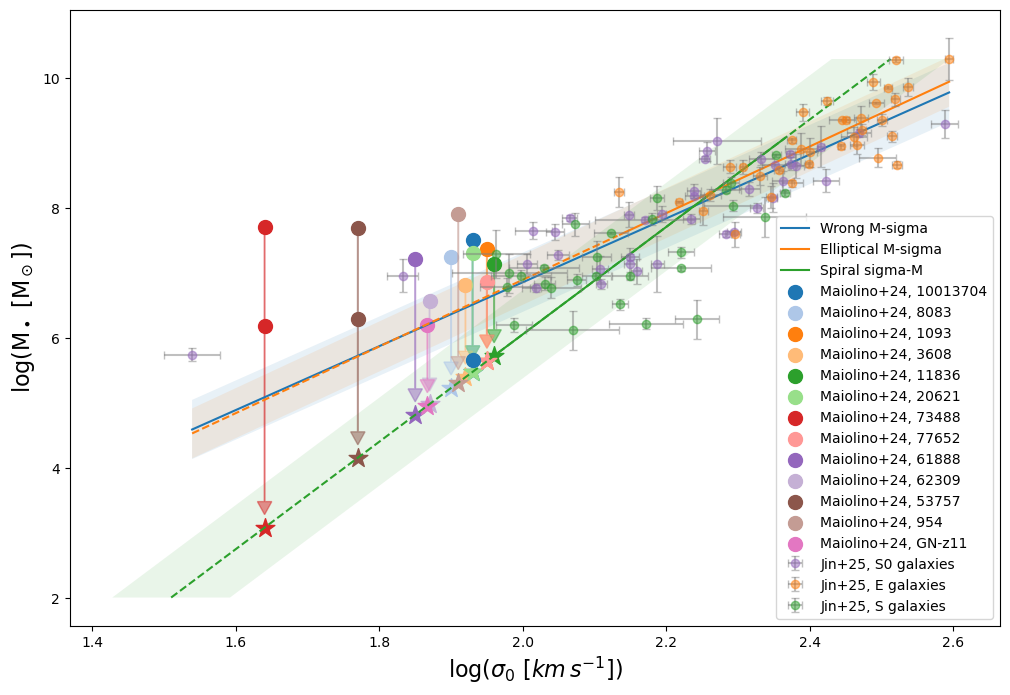}
    \caption{
    Linear regressions to the $z=0$ observational data from \citetalias{Jin:2025}, along with the published and revised high-$z$ black hole masses in 13 star-forming, late-type systems from \citet{Maiolino:2024}.
    The 101 galaxies are divided into 38 lenticular/S0 (\textcolor{S0}{$\bullet$}), 35 elliptical/E (\textcolor{E}{$\bullet$}), and 28 spiral/S (\textcolor{S}{$\bullet$}) galaxies.
    The \textcolor{matplotlib_blue}{{\hdashrule[0.35ex]{8mm}{1pt}{}}} line is the ``wrong'' fit to all 101 galaxies defined by Equation~\ref{eqn:sigma-wrong}.
    The \textcolor{matplotlib_orange}{{\hdashrule[0.35ex]{8mm}{1pt}{}}} line is the fit to the 35 elliptical (quenched) galaxies, according to the $M_\bullet\rightarrow\sigma_0$ causal direction, defined by Equation~\ref{eqn:elliptical-sigma}.
    The \textcolor{matplotlib_green}{{\hdashrule[0.35ex]{8mm}{1pt}{}}} line is the fit to the 28 spiral (star-forming) galaxies, according to the $\sigma_0\rightarrow M_\bullet$ causal direction, defined by Equation~\ref{eqn:spiral-inverted}.
    Each line is surrounded by its color's shaded region, depicting the intrinsic scatter bounds.
    The larger assorted color $\bullet$ points represent the published black hole masses from \citet{Maiolino:2024} with down arrows pointing to similar-colored $\bigstar$ stars, depicting our revised black hole masses.
    $N\!\!B$: three of the galaxies from \citet{Maiolino:2024} have dual BLR AGNs and are depicted here with two different $\bullet$ points each.
    The solid portions of the fitted relations indicate the range of $\sigma_0$ spanned by the corresponding observational samples used in the regression, while dashed extensions denote extrapolation beyond this range.
    In particular, the spiral (star-forming) galaxy relation is calibrated for $92\,\mathrm{km\,s^{-1}} \leq \sigma_0 \leq 232\,\mathrm{km\,s^{-1}}$.}
    \label{fig:regressions}
\end{figure}

In Table~\ref{tab:correct}, we present the published black hole masses ($\mathcal{M}_\bullet^\prime$) of \citet{Maiolino:2024} and their black hole to galaxy stellar mass ratios ($\mathcal{M}_\bullet^\prime/\mathcal{M}^\ast$), then we present our revised causally correct black hole mass estimates ($\mathcal{M}_\bullet$) and similarly revised black hole to galaxy stellar mass ratios ($\mathcal{M}_\bullet/\mathcal{M}^\ast$).\footnote{The black hole masses listed in Table~\ref{tab:correct} (Col.\ 8) are point estimates derived from the best-fit causally-informed relations; rigorous predictive uncertainties for individual objects require Monte Carlo sampling of the full regression posterior and are beyond the scope of this illustrative application.}
Figure~\ref{fig:regressions} illustrates the revised black hole masses on the $M_\bullet$--$\sigma_0$ diagram.
The figure further shows the graphical representation of our linear regressions for the fit to all galaxy types (Equation~\ref{eqn:sigma-wrong}), elliptical (quenched) galaxies (Equation~\ref{eqn:elliptical-sigma}), and spiral (star-forming) galaxies (Equation~\ref{eqn:spiral-inverted}).
This comparison shows a clear difference between the fits to star-forming versus quenched galaxies.
Moreover, this shows that a na\"{i}ve fit to all galaxies is similar to the fit for quenched galaxies, and produces overmassive black holes when applied to star-forming galaxies.

\begin{deluxetable*}{rcrllllll}
\tablecolumns{9}
\tablecaption{13 Revised High-$z$ Black Hole Masses from \citet{Maiolino:2024}}\label{tab:correct}
\tablehead{
\colhead{ID} & \colhead{Comp.} & \colhead{$z$} & \colhead{$\mathcal{M}^\ast$} & \colhead{$\mathcal{S}_0$} & \colhead{$\mathcal{M}_\bullet^\prime$} & \colhead{$\mathcal{M}_\bullet^\prime/\mathcal{M}^\ast$} & \colhead{$\mathcal{M}_\bullet$} & \colhead{$\mathcal{M}_\bullet/\mathcal{M}^\ast$}\\
\colhead{} & \colhead{} & \colhead{} & \colhead{[$\log\mathrm{M}_\sun$]} & \colhead{[$\log\mathrm{km/s}$]} & \colhead{[$\log\mathrm{M}_\sun$]} & \colhead{[dex]} & \colhead{[$\log\mathrm{M}_\sun$]} & \colhead{[dex]}\\
\colhead{(1)} & \colhead{(2)} & \colhead{(3)} & \colhead{(4)} & \colhead{(5)} & \colhead{(6)} & \colhead{(7)} & \colhead{(8)} & \colhead{(9)}
}
\startdata
\multirow{2}{*}{10013704} & BLR1 & \multirow{2}{*}{5.9} & \multirow{2}{*}{\phantom{1}$8.88\pm0.66$} & \multirow{2}{*}{$1.93^{+0.05}_{-0.06}$} & $5.65\pm0.31$ & $-3.23\pm0.73$ & \multirow{2}{*}{$5.47\pm0.83$} & \multirow{2}{*}{$-3.41\pm1.06$} \\
 & BLR2 & & & & $7.50\pm0.31$ & $-1.38\pm0.73$ & &  \\
8083 & & 4.6 & \phantom{1}$8.45\pm0.03$ & $1.90^{+0.06}_{-0.07}$ & $7.25\pm0.31$ & $-1.20\pm0.31$ & $5.23\pm0.88$ & $-3.22\pm0.88$ \\
1093 & & 5.6 & \phantom{1}$8.34\pm0.20$ & $1.95^{+0.05}_{-0.06}$ & $7.36^{+0.32}_{-0.31}$ & $-0.98\pm0.37$ & $5.64\pm0.83$ & $-2.70\pm0.85$ \\
3608 & & 5.3 & \phantom{1}$8.38^{+0.11}_{-0.15}$ & $1.92^{+0.06}_{-0.07}$ & $6.82^{+0.38}_{-0.33}$ & $-1.56\pm0.38$ & $5.39\pm0.88$ & $-2.99\pm0.89$ \\
11836 & & 4.4 & \phantom{1}$7.79\pm0.30$ & $1.96^{+0.05}_{-0.06}$ & $7.13\pm0.31$ & $-0.66\pm0.43$ & $5.72\pm0.83$ & $-2.07\pm0.88$ \\
20621 & & 4.7 & \phantom{1}$8.06\pm0.70$ & $1.93^{+0.06}_{-0.07}$ & $7.30\pm0.31$ & $-0.76\pm0.77$ & $5.47\pm0.88$ & $-2.59\pm1.12$ \\
\multirow{2}{*}{73488} & BLR1 & \multirow{2}{*}{4.1} & \multirow{2}{*}{\phantom{1}$9.78\pm0.20$} & \multirow{2}{*}{$1.64^{+0.11}_{-0.15}$} & $6.18\pm0.30$ & $-3.60\pm0.36$ & \multirow{2}{*}{$3.08\pm1.30$} & \multirow{2}{*}{$-6.70\pm1.32$} \\
 & BLR2 & & & & $7.71\pm0.30$ & $-2.07\pm0.36$ &  &  \\
77652 & & 5.2 & \phantom{1}$7.87^{+0.16}_{-0.28}$ & $1.95^{+0.06}_{-0.07}$ & $6.86^{+0.35}_{-0.34}$ & $-1.01\pm0.41$ & $5.64\pm0.88$ & $-2.23\pm0.91$ \\
61888 & & 5.9 & \phantom{1}$8.11\pm0.92$ & $1.85^{+0.07}_{-0.09}$ & $7.22\pm0.31$ & $-0.89\pm0.97$ & $4.81\pm0.97$ & $-3.30\pm1.33$ \\
62309 & & 5.2 & \phantom{1}$8.12^{+0.12}_{-0.13}$ & $1.87^{+0.07}_{-0.08}$ & $6.56^{+0.32}_{-0.31}$ & $-1.56\pm0.34$ & $4.98\pm0.94$ & $-3.14\pm0.94$ \\
\multirow{2}{*}{53757} & BLR1 & \multirow{2}{*}{4.4} & \multirow{2}{*}{$10.18^{+0.13}_{-0.12}$} & \multirow{2}{*}{$1.77^{+0.09}_{-0.11}$} & $6.29^{+0.33}_{-0.32}$ & $-3.89\pm0.35$ & \multirow{2}{*}{$4.15\pm1.09$} & \multirow{2}{*}{$-6.03\pm1.10$} \\
 & BLR2 & & & & $7.69^{+0.32}_{-0.31}$ & $-2.49\pm0.34$ &  &  \\
954 & & 6.8 & $10.66^{+0.09}_{-0.10}$ & $1.91\pm0.06$ & $7.90\pm0.30$ & $-2.76\pm0.31$ & $5.31\pm0.86$ & $-5.35\pm0.86$ \\
GN-z11 & & 10.6 & \phantom{1}$8.90^{+0.20}_{-0.30}$ & \nodata & $6.20\pm0.30$ & $-2.70\pm0.39$ & $4.95\pm1.16$\tablenotemark{$\ast$} & $-3.95\pm1.18$ \\
\enddata
\tablecomments{
Column~(1): Galaxy identification number.
Column~(2): Different components for galaxies with dual BLR AGNs.
Column~(3): Redshift.
Column~(4): Galaxy stellar mass (logarithmic solar masses).
Column~(5): Central stellar velocity dispersion (logarithmic km/s) approximated from the gaseous stellar velocity dispersion via corrections from \citet{Bezanson:2018}.
Column~(6): Black hole mass (logarithmic solar masses) estimated from local virial relations \citep{Reines:2013,Reines:2015}.
Column~(7): Black hole to galaxy stellar mass ratio (dex) according to \citet{Maiolino:2024}.
Column~(8): Black hole mass (logarithmic solar masses) estimated from Equation~\ref{eqn:spiral-inverted}.
Column~(9): Black hole mass to galaxy stellar mass ratio (dex) according to this work.
\tablenotetext{^\ast}{Derived from Equation~\ref{eqn:sigma-correct}.}
}
\end{deluxetable*}

We note that the observational samples used to calibrate the $M_\bullet$--$\sigma_0$ relations have a finite dynamic range in $\sigma_0$.
In particular, the sample of observed spiral galaxies used to derive the star-forming relation spans $92\,\mathrm{km\,s^{-1}} \leq \sigma_0 \leq 232\,\mathrm{km\,s^{-1}}$, whereas several of the high-redshift systems considered here have substantially lower inferred velocity dispersions.
Application of the relation in this regime therefore constitutes an extrapolation beyond the directly calibrated range.
This extrapolation is physically motivated by our causal analysis, which indicates that in star-forming systems the SMBH mass is the antecedent variable and $\sigma_0$ responds to SMBH growth, rather than the converse.
Under this causal interpretation, extending the relation toward lower $\sigma_0$ does not assume that the same dynamical processes operate unchanged, but instead reflects the expectation that lower-mass, younger systems occupy the same underlying evolutionary sequence.
Nevertheless, we emphasize that black hole mass estimates derived in this extrapolated regime should be regarded as approximate and illustrative, and not as precise predictions.
The dashed portions of the relations in Fig.~\ref{fig:regressions} explicitly indicate this extrapolation.

\begin{figure}
    \includegraphics[clip=true, trim= 3mm 3mm 3mm 10mm, width=\linewidth]{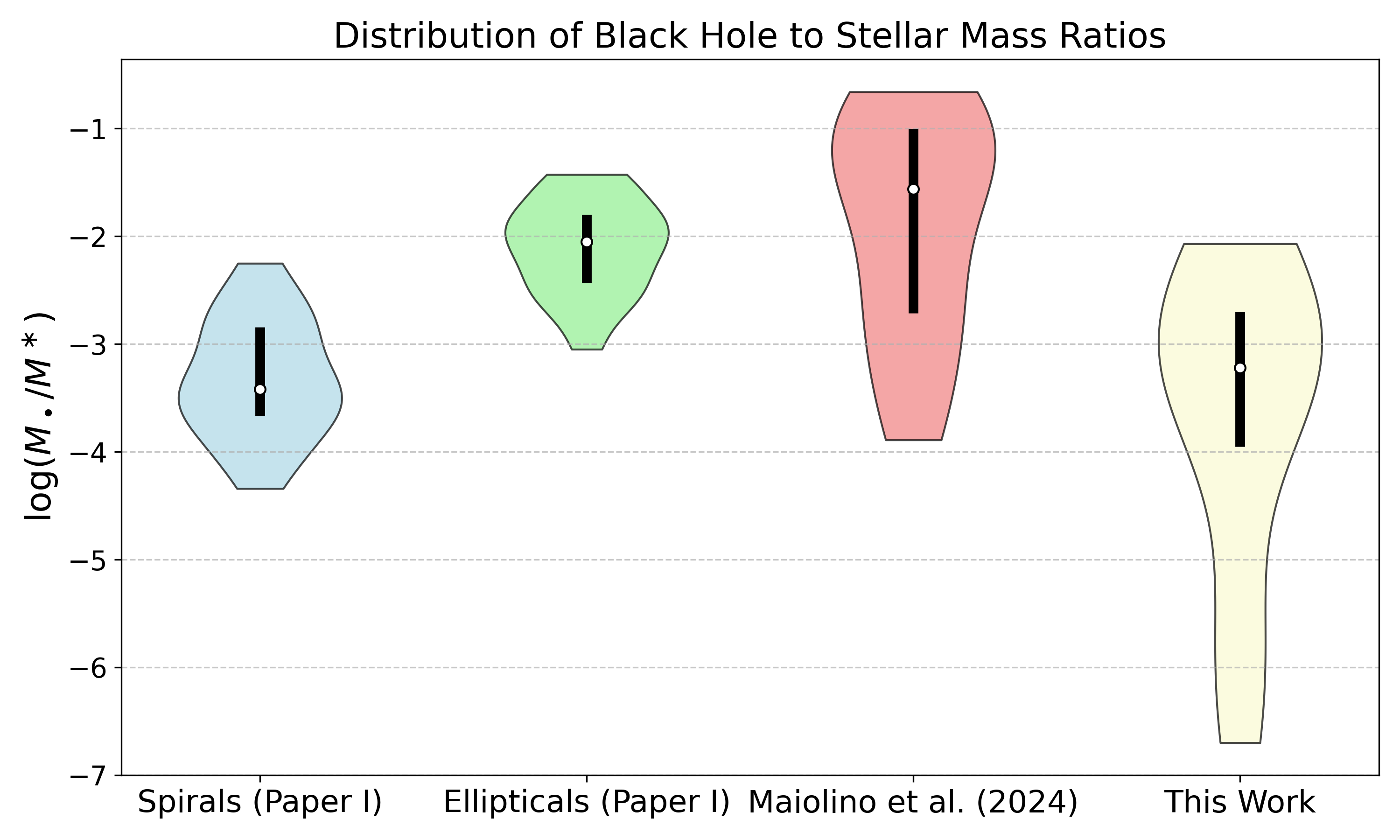}
    \caption{
    Violins plots showing the distribution of black hole to galaxy stellar mass ratios with indicators for their medians and the interquartile ranges.
    Moving from left to right: we show the sample of spiral galaxies from \citetalias{Jin:2025} with a median of $-3.42\pm0.44$\,dex, the elliptical galaxies from \citetalias{Jin:2025} with a median of $-2.05\pm0.29$\,dex, the high-$z$ late-type galaxies with black hole masses estimated by \citet{Maiolino:2024} with a median of $-1.56\pm0.74$\,dex, and the same high-$z$ late-type galaxies with black hole masses redetermined via our causally-informed correction (Equation~\ref{eqn:spiral-inverted}) with a median of $-3.22\pm0.63$\,dex.
    }
    \label{fig:violins}
\end{figure}

Our results show a clear decrease in the estimated black hole masses (and black hole to galaxy stellar mass ratios) for these high-redshift objects, with a median reduction of $1.93\pm0.51$\,dex.
Similarly, \citet{Rusakov:2025} measure SMBH masses for a population of LRDs that are two orders of magnitude lower than previous estimates, making them consistent with the observed SMBH--galaxy stellar mass relation at low redshifts.
This reduction effectively erases the appearance of being overmassive with high black hole to galaxy stellar mass ratios that resemble highly-evolved galaxies.
For instance, the median $\mathcal{M}_\bullet/\mathcal{M}^\ast$ ratios from \citetalias{Jin:2025} are $-2.05\pm0.29$\,dex and $-3.42\pm0.44$\,dex for ellipticals and spirals, respectively.
For comparison, the median $\mathcal{M}_\bullet^\prime/\mathcal{M}^\ast$ ratio from \citet{Maiolino:2024} is $-1.56\pm0.74$\,dex versus the median $\mathcal{M}_\bullet/\mathcal{M}^\ast$ ratio of $-3.22\pm0.63$\,dex after our reduction from causal correction (see Figure~\ref{fig:violins} for the distribution of ratios).
Therefore, the ratios were previously consistent with those of highly-evolved elliptical galaxies, whereas after our causal rectification the ratios now closely match the ratios of young star-forming galaxies, which should be the case for galaxies at $4.1<z<10.6$ without much cosmic time to evolve.
Indeed, this range of redshifts implies an age of the Universe at only 0.44--1.5\,Gyr after the Big Bang \citep[\hspace{-2.5mm}][]{Planck:2020}, leaving very little time for evolution.

\subsection{Conclusions}

Drawing together the threads of this investigation, we have successfully leveraged the burgeoning field of causal discovery, applied to cosmological hydrodynamical simulations from the NIHAO suite, to temporally resolve the causal relationship between SMBHs and their host galaxies.
Our simulations not only reproduce the observed causal dichotomy between spiral and elliptical galaxies---where SMBHs appear to drive evolution in star-forming spirals but are passive participants in ellipticals---but critically, they provide robust insight into its origin.
The key finding reveals a distinct causal reversal coincident with the epoch of peak star-formation, transitioning from SMBH-driven influence during active star-formation phases to a more passively correlated growth governed by hierarchical assembly in quenched systems.
These results offer compelling support for theoretical models positing that AGNs feedback plays a crucial role in shaping star-forming galaxies, while subsequent growth is primarily dictated by the host galaxy's merger and accretion history.
This study underscores the immense potential of causal discovery techniques in unraveling the complex interplay governing not only SMBH and their host galaxies, but in any set of intertwined properties in other fields of astrophysics \citep[e.g.,][]{Jin:2025b,Davis:2025a,Davis:2025b,Zhang:2025b,Desmond:2025}.
Finally, a causally-reversed $M_\bullet$--$\sigma_0$ relation casts considerable doubt on the validity of SMBH mass estimates for distant galaxies and challenges the prevailing view of overmassive black holes existing in the early Universe.
Toward that problem, we offer a set of causally-informed relations to produce black hole mass estimates with improved fidelity that are consistent with $z=0$ SMBH--galaxy stellar mass ratios.

\begin{acknowledgments}
This material is based on work supported by Tamkeen under the NYU Abu Dhabi Research Institute grant CASS.
F.Y.\ is supported by the NSF of China (grants 12133008, 12192220, 12192223, and 12361161601).
This research has used NASA's Astrophysics Data System.
This research was carried out on the high-performance computing resources at New York University Abu Dhabi.
\end{acknowledgments}

\software{
\href{https://causal-learn.readthedocs.io/en/latest/}{\textcolor{linkcolor}{\texttt{causal-learn}}} \citep{causallearn},
\href{https://cosmocalc.icrar.org/}{\textcolor{linkcolor}{\texttt{Cosmology Calculator}}},
\href{https://www.gymlibrary.dev/index.html}{\textcolor{linkcolor}{\texttt{Gym}}} \citep{gym},
\href{https://github.com/CullanHowlett/HyperFit}{\textcolor{linkcolor}{\texttt{Hyper-Fit}}} \citep{Robotham:2015,Robotham:2016},
\href{https://jax.readthedocs.io/en/latest/}{\textcolor{linkcolor}{\texttt{JAX}}} \citep{jax2018github},
\href{https://github.com/matplotlib/matplotlib}{\textcolor{linkcolor}{\texttt{Matplotlib}}} \citep{Hunter:2007},
\href{https://networkx.org/}{\textcolor{linkcolor}
{\texttt{NetworkX}}} \citep{networkx},
\href{https://github.com/numpy/numpy}{\textcolor{linkcolor}{\texttt{NumPy}}} \citep{harris2020array},
\href{https://pandas.pydata.org/}{\textcolor{linkcolor}{\texttt{Pandas}}} \citep{McKinney_2010},
\href{https://pgmpy.org/}{\textcolor{linkcolor}{\texttt{pgmpy}}} \citep{ankan2015pgmpy},
\href{https://pygraphviz.github.io/}{\textcolor{linkcolor}{\texttt{PyGraphviz}}},
\href{https://pynbody.readthedocs.io/latest/}{\textcolor{linkcolor}{\texttt{pynbody}}} \citep{pynbody},
\href{https://www.python.org/}{\textcolor{linkcolor}{\texttt{Python}}} \citep{Python},
\href{https://github.com/scipy/scipy}{\textcolor{linkcolor}{\texttt{SciPy}}} \citep{Virtanen_2020},
\href{https://seaborn.pydata.org/}{\textcolor{linkcolor}{\texttt{seaborn}}} \citep{Waskom2021}
}

\section*{ORCID iDs}

\begin{CJK*}{UTF8}{gbsn}
\begin{flushleft}
Benjamin L.\ Davis \scalerel*{\includegraphics{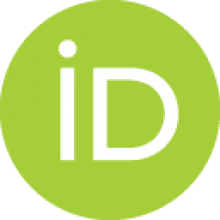}}{B} \url{https://orcid.org/0000-0002-4306-5950}\\
Zehao Jin (金泽灏) \scalerel*{\includegraphics{orcid-ID.png}}{B} \url{https://orcid.org/0009-0000-2506-6645}\\
Mario Pasquato \scalerel*{\includegraphics{orcid-ID.png}}{B} \url{https://orcid.org/0000-0003-3784-5245}\\
Andrea Valerio Macci\`{o} \scalerel*{\includegraphics{orcid-ID.png}}{B} \url{https://orcid.org/0000-0002-8171-6507}\\
Feng Yuan (袁峰) \scalerel*{\includegraphics{orcid-ID.png}}{B} \url{https://orcid.org/0000-0003-3564-6437}
\end{flushleft}
\end{CJK*}

\bibliography{bibliography}

\end{document}